\documentclass[acmsmall,nonacm]{acmart}

\setcitestyle{nosort}
\AtEndPreamble{\theoremstyle{acmdefinition}
    }

\usepackage{amsmath}
\usepackage{mathrsfs}
\usepackage{tikz}
\let\rel\mathbf
\let\clo\mathscr
\let\tup\mathbf

\let\equals\approx
\let\epsilon\varepsilon

\DeclareMathOperator{\Pol}{Pol}

\DeclareMathOperator{\sel}{sel}
\DeclareMathOperator{\CSP}{CSP}
\DeclareMathOperator{\PCSP}{PCSP}
\newcommand{\NP}{\text{NP}}
\newcommand{\Ptime}{\text{P}}

\newcommand{\Sat}{\text{\scshape Sat}}
\newcommand{\nae}{\text{\upshape NAE}}
\newcommand{\NAE}{\rel H}

\DeclareMathOperator{\ar}{ar}
\DeclareMathOperator{\id}{id}

\newcommand{\yes}{{\scshape yes}}
\newcommand{\no}{{\scshape no}}
\DeclareMathOperator{\BLP}{BLP}

\DeclareMathOperator{\AIP}{AIP}

\newcommand{\conv}{{\text{conv}}}
\newcommand{\aff}{{\text{aff}}}

\makeatletter
\let\@authorsaddresses\@empty
\makeatother

\title{An Invitation to the Promise Constraint Satisfaction Problem}

\author{Andrei Krokhin}
\affiliation{\institution{Durham University}
  \city{Durham}
  \country{UK}
}

\author{Jakub Opršal}
\affiliation{\institution{Institute of Science and Technology Austria}
  \city{Klosterneuburg}
  \country{Austria}
}

\begin{abstract}
  The study of the complexity of the constraint satisfaction problem (CSP), centred around the Feder-Vardi Dichotomy Conjecture, has been very prominent in the last two decades. After a long concerted effort and many partial results, the Dichotomy Conjecture has been proved in 2017 independently by Bulatov and Zhuk.
  At about the same time, a vast generalisation of CSP, called promise CSP, has started to gain prominence. In this survey, we explain the importance of promise CSP and highlight many new very interesting features that the study of promise CSP has brought to light. The complexity classification quest for the promise CSP is wide open, and we argue that, despite the promise CSP being more general, this quest is rather more accessible to a wide range of researchers than the dichotomy-led study of the CSP has been.
\end{abstract}

\thanks{This survey appeared in ACM SIGLOG News, Volume 9, Issue 3, July 2022, pp~30–59, \href{https://doi.org/10.1145/3559736.3559740}{doi:10.1145/3559736.3559740}.}

\begin{document}

\maketitle

\section{Introduction}
  \label{sec:intro}

What kind of inherent mathematical structure makes a~computational problem \emph{tractable}, i.e., polynomial-time solvable (assuming $\Ptime\ne\NP$)? Finding a~general answer to this question is one of the fundamental goals of theoretical computer science. The \emph{constraint satisfaction problem (CSP)} and its variants are extensively used towards this ambitious goal for two reasons: on the one hand, the CSP framework is very general and includes a~wide variety of computational problems, and on the other hand, this framework has very rich mathematical structure providing an excellent laboratory both for complexity classification methods and for algorithmic techniques.

In this section we informally discuss the CSP and the \emph{promise CSP (PCSP)}, which is the main subject of this survey. Formal definitions and more technical aspects can be found in subsequent sections.

The (decision version of the) CSP can be defined in several equivalent ways. One way is to define it as the problem of deciding the existence of a homomorphism from one finite relational structure (for example, a digraph) to another. We focus on the so-called non-uniform CSP --- when the target structure $\rel A$ is fixed (and is often called a \emph{template} or a \emph{constraint language}). The problem, denoted $\CSP(\rel A)$, becomes the problem of deciding whether an input structure $\rel I$ homomorphically maps to $\rel A$ (we write $\rel I\rightarrow \rel A$). Problems of this form include $k$-{\Sat}, graph $k$-colouring, and solving systems of linear equations over a fixed finite field. 

A~major line of research in CSP focuses on identifying tractable cases and understanding the mathematical structure enabling tractability \citep[see][]{KZ17}.
\citet{FV98} conjectured that for each relational structure $\rel A$, the problem $\CSP(\rel A)$ is either tractable (i.e., in P), or NP-complete, i.e., it cannot be \NP-intermediate. Due to its very impressive scope, this CSP Dichotomy Conjecture has been widely considered one of the most important open problems in theoretical computer science. It inspired a~very active research programme in the last 25 years, which culminated in two independent proofs of the conjecture in 2017, one by \citet{Bul17} and the other by \citet{Zhu17,Zhu20}. The key part of both proofs are polynomial-time algorithms for the tractable cases, which heavily exploit structural properties of finite universal algebras associated with instances of the problem.

Approximation provides a natural way of coping with NP-hardness.  One natural approximation version of the graph $k$-colouring problem is the \emph{approximate graph colouring} problem, where one needs to find a $c$-colouring for a given $k$-colourable graph, for $c \ge k$. This problem is very stubborn --- it is believed to be \NP-hard for all constants $3\le k \le c$, but, despite being studied for more than 45 years~\cite{GJ76}, the progress towards proving this has been very limited. Even for $k=3$, the best \NP-hardness result (without making additional assumptions) has recently been improved from $c=4$ \cite*{KLS00,GK04} to $c=5$ \cite*{BKO19, BBKO21}, while the best (in terms of $c$) polynomial-time algorithm for colouring a 3-colourable graph with $n$ vertices uses about $n^{0.199}$ colours \cite{KT17}.
The huge gap between best known negative and positive results shows that we are far from understanding the mathematical nature of this problem. We note that, by additionally assuming different (perfect-completeness) variants of the Unique Games Conjecture, NP-hardness of all approximate graph colouring problems (with $k\ge 3$) was proved in \cite{BKLM22,DMR09,GS20}, but further in this survey we focus on results that do not make any complexity assumptions other than $\Ptime\ne \NP$.

Given the success of the research programme aimed at the CSP Dichotomy Conjecture, it was suggested in \cite*{AGH17,BG16,BG21} that perhaps progress on approximate graph colouring and similar problems can be made by looking at a more general picture, using the CSP paradigm. This naturally led to the following formulation: fix not just one relational structure $\rel A$, but also another structure $\rel B$ such that $\rel A\rightarrow \rel B$. Then, given an input structure $\rel I$ with a promise that $\rel I\rightarrow \rel A$, the goal is to \emph{find} a homomorphism from $\rel I$ to $\rel B$ (which must exist because $\rel I\rightarrow\rel A\rightarrow\rel B$). Note that a homomorphism to $\rel A$ is promised to exist, but is not given as a part of input. This promise problem is denoted $\PCSP(\rel A,\rel B)$ and is called a \emph{promise CSP}.  When $\rel A=\rel K_k$ and $\rel B=\rel K_c$ are the complete graphs on $k$ and $c$ (resp.) vertices, $\PCSP(\rel A,\rel B)$ is precisely the approximate graph colouring problem.
The assumption $\rel A\rightarrow \rel B$ is necessary for the problem to make sense.
The problem defined above is a search problem. The standard \emph{decision} problem associated with it is to distinguish input structures $\rel I$ such that $\rel I \rightarrow \rel A$ from those satisfying $\rel I\not\rightarrow \rel B$.

Let us make a few easy, but important, observations:
\begin{itemize}
  \item When $\rel A = \rel B$, the decision version of the problem $\PCSP(\rel A, \rel A)$ is precisely $\CSP(\rel A)$. Thus the PCSP framework generalises the CSP framework.
  \item The decision version is always reducible to the search version (simply run the search algorithm and then verify its output --- if the algorithm crashes or the output is wrong, then we cannot have $\rel I\rightarrow\rel A$), but it is open whether the two versions are always equivalent.
  \item Both decision and search $\PCSP(\rel A,\rel B)$ are in $\Ptime$ if at least one of $\CSP(\rel A)$ and $\CSP(\rel B)$ is in $\Ptime$. This is because tractable decision CSPs have a tractable search problem, as can be shown by a simple self-reduction trick \cite[see, e.g.,][]{BJK05}.
\end{itemize}

Problems $\PCSP(\rel A, \rel B)$ can be viewed as approximation versions of $\CSP(\rel A)$ on satisfiable instances, where approximation is understood in a qualitative way (i.e., relaxing constraints, as opposed to allowing to violate some of the constraints).  The quest to classify the complexity of problems of the form $\PCSP(\rel A,\rel B)$ can also be seen the following way. Fix a structure $\rel A$ such that $\CSP(\rel A)$ is \NP-complete. What are the structures $\rel B$ such that $\PCSP(\rel A, \rel B)$ is in \Ptime? What are the structures $\rel B$ such that $\PCSP(\rel A, \rel B)$ is in \NP-hard? What mathematical properties of the structures determine the complexity?

A different way to view $\PCSP(\rel A, \rel B)$ is as $\CSP(\rel B)$ with restricted inputs (for decision problems, we can view it as either $\CSP(\rel A)$ or $\CSP(\rel B)$ restricted to inputs where the two answers agree). This means that not only the classification problem with respect to P or NP-hardness, but many other questions that have been studied for CSPs (e.g., classification with respect to other complexity classes or with respect to definability in various logics) make perfect sense for PCSPs, possibly with a minor adjustment.

The so-called algebraic approach was the key tool in attacking the CSP Dichotomy Conjecture and related problems. This approach has two layers. The first layer is the framework of polymorphisms. They are simple combinatorial objects --- for example, a polymorphism of a graph $\rel G$ is simply a homomorphism from a direct power of $\rel G$ to $\rel G$. Polymorphisms are known to control gadget reductions between CSPs (as we discuss in detail later). The second, much more advanced, layer is the structural theory of universal algebra. A complete translation of the CSP Dichotomy Conjecture into universal algebra allowed the whole universal algebraic community to contribute to the progress on the conjecture, which eventually lead to enormous progress both in CSP and in universal algebra itself. This theory suited the dichotomy problem so well that it enabled fantastic progress, but, on the other hand, it became so dominant in the theory of CSP that it possibly prevented some mathematical ideas from being developed and some researchers from trying to work in this direction.
We will try to argue that the study of PCSP can facilitate new ideas and connections that have been overlooked so far.

The structural theory of universal algebras does not seem to be applicable for PCSPs, but polymorphisms still are. Say, for graphs, they can now be thought of as homomorphisms from a direct power of one graph $\rel G$ to another graph $\rel H$. So, one needs to get more creative in using polymorphisms. Roughly, there are three general ways to use polymorphisms for PCSPs (which we discuss in detail in Sections \ref{sec:tract}--\ref{sec:hardness}, respectively):
\begin{itemize}
  \item When polymorphisms of a PCSP are rich enough, they can be used as rounding procedures in polynomial-time algorithms. One relaxes a given PCSP instance in some way, solves the relaxation, and then rounds the relaxed solution by applying appropriate (for the relaxation) polymorphisms. 
  \item Polymorphisms provide a convenient tool for characterising the existence of specific types of reductions between PCSPs. That is, one PCSP reduces to another PCSP by a specific type of reduction (say, a gadget reduction) if and only if the polymorphisms of the involved PCSPs are related in a certain way.
  \item If polymorphisms of a PCSP are limited enough, this can be used directly to construct NP-hardness proofs for the PCSP.
\end{itemize}
The three general guidelines described above are far from being fully formed. There is much to be developed here: what exactly ``rich enough'' and ``limited enough'' should mean (and whether there is a gap between them), what useful types of reductions can characterised in this way, and so on. 
     
The study of PCSPs brought to light many new aspects that were not present in the study of CSPs. Here we give a (non-exhaustive) list of such aspects.
\begin{itemize}
  \item The dichotomy-led study of CSPs sometimes received criticism for not including specific problems of wide interest whose complexity was open. The extension to PCSP includes such problems, the most prominent one being approximate graph colouring.
  \item When the dichotomy-led study of CSPs started in the 1990s, the Feder-Vardi conjecture was supported by two important special cases that were fully classified by then: the Boolean case \cite{Sch78} and the graph homomorphism case (also known as $H$-colouring) \cite{HN90}. 
    For PCSPs, even these two cases (i.e., when both $\rel A$ and $\rel B$ are two-element structures and when both $\rel A$ and $\rel B$ are undirected graphs, respectively) are currently wide open. In particular, as tempting as it might be, we cannot now conjecture a dichotomy for PCSPs because we simply do not have enough evidence.
  \item Already after a few years of research, unexpected connections with new areas appeared, such as the connection with algebraic topology \cite*{KOWZ20}, which appears to play a significant role in approximate graph colouring and probably beyond. Other examples include matrix analysis \cite{CZ22} and Boolean function analysis \cite*{BGS21}.
  There is a clear feeling that more new connections will show up because the study of PCSPs seems to be open to contributions from many different communities.
  \item Already on the existing evidence, new reductions and new algorithms that were overlooked in the study of CSPs can become quite important. This can possibly lead to a new simpler proof of the Feder-Vardi conjecture.
  \item For each CSP, the decision and the search problems are always equivalent \cite{BJK05}. For PCSPs, there is always a reduction from decision to search, but the other direction is open. In fact, the design of efficient search algorithms for PCSPs appears to bring brand new challenges \cite[see, e.g.,][]{BGWZ20}.
  \item The study of infinite-domain CSP (i.e., $\CSP(\rel A)$ when the template $\rel A$ is infinite) has been developing over the last 20 years \cite{Bod21}, using the algebraic approach as well as model theory and Ramsey theory. The transfer of methods and results between the infinite-domain and finite-domain cases has so far been mainly from latter to the former. 
    The framework of (finite-domain) PCSPs appears to require transfer in the other direction too.
\end{itemize}

\section{Basic notions and examples}
  \label{sec:prelim}

In this section, we explain the basics of the theory of promise CSPs and show a few key examples. Throughout the paper, we write $[n]$ for the set $\{1,2,\ldots, n\}$.

\subsection{CSPs and PCSPs}

We define CSP as a homomorphism problem for finite relational structures.

\begin{definition}
A \emph{(relational) structure} is a tuple $\rel A = (A; R_1^{\rel A},\ldots,R_l^{\rel A})$ where each $R_i^{\rel A}\subseteq A^{\ar(R_i)}$ is a relation on $A$ of arity $\ar(R_i)\ge 1$. We say that $\rel A$ is finite if its \emph{domain} $A$ is finite. We assume that structures are finite, unless specified otherwise.
Two structures $\rel A$ and $\rel B$ are called \emph{similar} if they have the same number of relations and the arities of corresponding relations coincide.
\end{definition}

For example, a~(directed) graph is relational structure with one binary relation. Any two graphs are similar structures.

We often use a~single letter instead of $R_i$ to denote a~relation of a~structure, e.g., $S^{\rel A}$ would denote a~relation of $\rel A$, the corresponding relation in a~similar structure $\rel B$, would be denoted by $S^{\rel B}$. Throughout the paper we denote the domains of structures $\rel A$, $\rel B$, $\rel K_n$ and so on by $A$, $B$, $K_n$, etc., respectively.

\begin{definition}
  For two similar relational structures $\rel A$ and $\rel B$, a~\emph{homomorphism} from $\rel A$ to $\rel B$ is a~map $h\colon A \to B$ such that, for each~$i$, if $(a_1,\ldots,a_{\ar(R_i)})\in R_i^{\rel A}$, then $(h(a_1),\ldots,h(a_{\ar(R_i)}))\in R_i^{\rel B}$.
  We write $h\colon \rel A\to \rel B$ to denote this, and simply $\rel A \rightarrow \rel B$ to denote that a homomorphism from $\rel A$ to $\rel B$ exists.  In the latter case, we also say that $\rel A$ maps homomorphically to $\rel B$.
\end{definition}

We defined $\CSP(\rel B)$ in the introduction as a homomorphism problem, i.e., given a structure $\rel I$ similar to $\rel B$, decide if $\rel I \to \rel B$.
There are two other standard ways to define $\CSP(\rel B)$, all are equivalent. One of them views the domain of $\rel I$ as a set of variables, the domain of $\rel B$ as possible values for these variables, and every tuple $\tup v$ in a relation $R^{\rel I}$  in $\rel I$ gives rise to a constraint $(\tup v, R^{\rel B})$, postulating that the tuple of values for $\tup v$ must be in the relation $R^{\rel B}$. The question is then whether there is an assignment of values to the variables that satisfies all constraints. We will often use this variables-constraints view alongside the homomorphism view.
The other way formalises CSP as the problem of satisfiability of primitive-positive (or pp-, for short) formulas in $\rel B$ \cite[for more details, see][]{BKW17}.

It is well-known and easy to see that if a $\CSP(\rel B)$ is NP-complete for some $\rel B$ (say, graph 3-colouring), there is no polynomial time algorithm that would \emph{find} a homomorphism $\rel I \to \rel B$ given the \emph{promise} that such a homomorphism exists (e.g., it is NP-hard to find a 3-colouring of a graph even when you know that it exists). The \emph{promise CSP} asks whether finding a homomorphism becomes easier if we relax our goal (e.g., we ask to colour a 3-colourable graph with 6 colours). Formally, promise CSP is defined as follows.

\begin{definition}
  Given two relational structures $\rel A$, $\rel B$ such that $\rel A \to \rel B$, the \emph{search} version of \emph{promise CSP} with template $(\rel A, \rel B)$ is, given an~input structure $\rel I$ that is promised to map homomorphically to $\rel A$, \emph{find} a~homomorphism $h\colon \rel I\to \rel B$.

  The \emph{decision promise CSP} with template $(\rel A, \rel B)$ as above is, given an input structure $\rel I$ similar to $\rel A$ and $\rel B$, output \yes{} if $\rel I \rightarrow \rel A$, and \no{} if $\rel I \not\rightarrow \rel B$. The promise here is that the input falls into one of the two categories.
\end{definition}

Both of the search and the decision versions are often denoted by the same symbol $\PCSP(\rel A, \rel B)$, and sometimes not so much attention is given to the distinction between them. Throughout the survey, we do not specify which version (decision or search) is being discussed whenever the results apply to both versions. When the distinction between search and decision is significant, we comment on it.

Let us give some examples of the problems of the form $\PCSP(\rel A,\rel B)$ which are proper promise problems, i.e., not of the form $\CSP(\rel A)$. Note that the interesting ones consists of templates $(\rel A,\rel B)$ where both $\CSP(\rel A)$ and $\CSP(\rel B)$ are NP-complete.

\begin{example}[$(2+\epsilon)$-\Sat] \label{ex:2+eps}
  $(2+\epsilon)$-{\Sat} is a collective name for a family of promise versions of $(2k+1)$-\Sat{} problems, where $k\ge 1$. Loosely speaking the goal is to find a satisfying assignment to a $(2k+1)$-\Sat{} instance being promised that there is an assignment that satisfies at least $k$ literals in each clause. Formally, the template is the following pair of relational structures:
  \begin{align*}
      \rel A_k &= (\{0,1\}; \{t \in \{0,1\}^{2k+1}\mid \operatorname{Ham}(t) \ge k\}, {\ne_2}),\\
      \rel B_k &= (\{0,1\}; \{t \in \{0,1\}^{2k+1}\mid \operatorname{Ham}(t)\ge 1\}, {\ne_2}).
  \end{align*}
  Here, $\operatorname{Ham}$ denotes the Hamming weight of the tuple, and $\ne_2$ the binary inequality relation on $\{0, 1\}$ (which is there to capture negative literals in clauses).
  The problems $\PCSP(\rel A_k,\rel B_k)$ were proved to be \NP-hard in \cite{AGH17}, where also the general framework of PCSPs and the name \emph{promise constraint satisfaction problem} were introduced.
\end{example}

\begin{example}[Promise 1-in-3-\Sat] \label{ex:1in3}
  One notable example of a promise CSP is the promise version of monotone 1-in-3-\Sat. Generally, the problem is given a satisfiable instance of monotone 1-in-3-\Sat, find a solution in some other structure $\rel B$ with a ternary (symmetric) relation $R^{\rel B}$. This is the problem $\PCSP(\rel T, \rel B)$ where
  \[
    \rel T = (\{0,1\}; \{(1,0,0),(0,1,0),(0,0,1)\}).
  \]
  What are the structures $\rel B$ for which $\PCSP(\rel T, \rel B)$ is tractable (respectively, NP-hard)? This question in this generality was posed in \cite{BBKO21}, and partially answered in \cite{BBB21}.

  Consider the case when $\rel B=\rel H_2$ is the Boolean ternary not-all-equal structure, i.e., $\rel H_2 = (\{0, 1\}; \{(x, y, z) \mid \text{not } (x = y = z)\})$. Then $\CSP(\rel H_2)$ is the Not-All-Equal-{\Sat} problem. $\PCSP(\rel T,\rel H_2)$ provides a range of interesting new features not found in CSPs: for example, even though $\CSP(\rel T)$ and $\CSP(\rel H_2)$ are (well-known) $\NP$-hard problems, the problem $\PCSP(\rel T, \rel H_2)$ was shown to be in $\Ptime$ in \cite{BG21}, along with a range of similar problems.
  We will discuss several other interesting properties of $\PCSP(\rel T, \rel H_2)$ later in this survey.
\end{example}

\begin{example}[Approximate graph colouring] \label{ex:approx-graph-col}
  Probably the most famous example of promise constraint satisfaction problem is the \emph{approximate graph colouring} that we already mentioned in the introduction.
  The problem is to find, for fixed $c \geq k$, a $c$-colouring of a given $k$-colourable graph. The template of this PCSP is a pair of cliques, $(\rel K_k, \rel K_c)$, where $\rel K_n = (\{0,\ldots,n-1\}; {\ne_n})$.
  The problem has been conjectured to be NP-hard for any fixed $3\le k\le c$, but this is still open in many cases. The current state-of-the-art for NP-hardness is $c = 2k - 1$ for $k = 3, 4, 5$ \cite{BBKO21} and $c = \binom{k}{\lfloor k/2\rfloor} - 1$ for $k\ge 5$ \cite{KOWZ20}. 
  Some unconditional results about the applicability of specific algorithms for approximate graph colouring can be found in \cite{AD22,CZ22a}.
\end{example}

\begin{example}[Approximate graph homomorphism] \label{ex:approx-graph-hom}
  Approximate graph homomorphism is a natural generalisation of approximate graph colouring, it is the PCSP where the template consists of two (undirected) graphs.
  We note that $\CSP(\rel H)$ for a~fixed (undirected) graph $\rel H$ is often called $\rel H$-colouring. A~well-known result by \citet{HN90} states that $\rel H$-colouring is solvable in polynomial time if $\rel H$ is bipartite or has a loop, and it is $\NP$-complete otherwise. \Citet{BG21} conjectured that the Hell-Nešetřil dichotomy extends to the promise problem, i.e., that $\PCSP(\rel G, \rel H)$ is \NP-hard for any non-bipartite loopless undirected $\rel G, \rel H$ with $\rel G \rightarrow\rel H$. It is not hard to see that it is sufficient to prove this conjecture for all cases where $\rel G$ is an odd cycle and $\rel H$ is a $c$-clique with $c\ge 3$. The case $\rel G = \rel C_{2k+1}$ (odd cycle), $\rel H = \rel K_3$ has been shown to be NP-hard in \cite{KOWZ20}. 
  If we consider directed graphs as well, then, extending a similar result for CSPs \cite{FV98}, it was shown in \cite{BG21} that every problem $\PCSP(\rel A, \rel B)$ is polynomial-time equivalent to $\PCSP(\rel G, \rel H)$ for suitable digraphs $\rel G, \rel H$.
\end{example}

\begin{example}[Approximate hypergraph colouring] \label{ex:hypergraph-colouring}
  This problem is similar to Example~\ref{ex:approx-graph-col}, but uses the ``not-all-equal'' relation
  \[
    \nae_k = \{0,\ldots,k-1\}^3\setminus \{(a,a,a)\mid a\in \{0,\ldots,k-1\}\}
  \]
  instead of $\neq_k$, and similarly for $c$, i.e., we are talking about $\PCSP(\NAE_k,\NAE_c)$ where
  \(
    \NAE_n = (\{0,\ldots,n-1\}; {\nae_n}).
  \)
  Note that $\rel H_2$ is the template of NAE-Sat as in Example~\ref{ex:1in3}.
  A colouring of a~hypergraph is an assignment of colours to its vertices that leaves no hyperedge monochromatic.  Thus, in (the search variant of) this problem one needs to find a $c$-colouring for a given $k$-colourable 3-uniform hypergraph.  This problem has been proved to be \NP-hard for any fixed $2\le k\le c$ \cite{DRS05}. Some variants of this problem that involve rainbow or strong versions of hypergraph colouring, have been studied \cite{ABP20, BG16, GL17, GS20a}, but full classifications are still open there.
\end{example}

\subsection{A theory of gadget reductions}
  \label{sec:gadgets}

We will briefly outline a general theory of promise CSPs, that is based on the influential `algebraic approach to CSP' pioneered by \citet{JCG97} and \citet{BJK05} (see \cite{BKW17} for a gentle introduction).
The core story of the algebraic approach to CSP is an abstraction from the concrete CSP template $\rel A$ to an algebraic structure, denoted by $\Pol(\rel A)$ and called \emph{polymorphisms of $\rel A$}, assigned to it. The main theorem then asserts that the complexity of $\CSP(\rel A)$ (up to log-space reductions) only depends on the algebraic properties of $\Pol(\rel A)$. The situation for promise CSP is similar: there is a natural generalisation of polymorphisms to the promise setting, but the polymorphisms do not form an \emph{algebra} or a \emph{clone}, but a weaker structure that is called a \emph{minion}. The core thesis is still valid — the complexity of $\PCSP(\rel A, \rel B)$ only depends on the properties of polymorphisms from $\rel A$ to $\rel B$, up to log-space reductions. We will define the notion of polymorphism and minion below together with a precise statement from which the above thesis immediately follows.

\subsubsection{Gadgets and the scope of the theory}
\label{sec:gadgets-scope}

Often the algebraic theory investigates certain constructions $\gamma$ (so-called \emph{pp-interpretations}, \emph{pp-constructions}, or \emph{pp-powers}, we refer to \cite[Section 3]{BKW17} for definitions of these notions) of the \emph{template} $\rel B$ that allow for efficient complexity reduction from $\CSP(\gamma(\rel B))$ to $\CSP(\rel B)$. Here, we would like to focus more on the transformation (of instances) used in the reduction, as opposed to transformation of the template. We discuss the connection with pp-powers near the end of this subsection.

Formally, by a \emph{reduction} from problem $A$ to problem $B$ we mean an efficiently computable function $\phi$ from instances of problem $A$ to instances of problem $B$ such that a solution to an instance $I$ can be reconstructed from a solution to the instance $\phi(I)$, e.g., if both $A$ and $B$ are promise decision problems, we require that $\phi$ maps {\yes} instances of $A$ to {\yes} instances of $B$ and {\no} instances of $A$ to {\no} instances of $B$. For search problems, we moreover require an efficiently computable function from solutions of $\phi(I)$ to solutions of $I$.

A gadget replacement (or local replacement) is a much used technique for constructing reductions in complexity theory, where each basic unit of an instance is replaced by a different structure in a uniform way.
Gadget reductions used in (promise) CSP are precisely this type of reductions, but with a more formal notion of a gadget.

We now give formal definitions of gadgets and gadget reductions that we will use. Let us remark that these definitions can be equally well expressed using pp-formulas.

For a gadget reduction from $\CSP(\rel A)$ to $\CSP(\rel B)$, we need the following data:
\begin{itemize}
  \item a structure $\rel G$ similar to $\rel B$, to be used as a gadget for variables, and
  \item a structure $\rel G_R$ for each ($k$-ary) relational symbol $R$ of $\rel A$, with fixed homomorphisms $e_{R,i}\colon \rel G \to \rel G_R$ ($1\le i\le k$), to be used as a gadget for constraints involving $R$.
\end{itemize}

\begin{definition}[Gadget replacement]
  Assuming the data as above, let $\rel I$ be an instance of $\CSP(\rel A)$, i.e., a structure similar to $\rel A$. We define a gadget replacement as the following construction applied to $\rel I$, resulting in a structure $\phi(\rel I)$ similar to $\rel B$.
  \begin{itemize}
    \item For each element $v\in I$, introduce into $\phi(\rel I)$ a copy of $\rel G$ denoted by $\rel G^v$. We denote the elements of $\rel G^v$ by $g_v$ where $g \in G$.
    \item For each constraint of $\rel I$, i.e., each relational symbol $R$ say of arity $k$ and all $(v_1, \dots, v_k)\in R^{\rel I}$, we introduce into $\phi(\rel I)$ a copy of $\rel G_R$ denoted by $\rel G_R^{v_1, \dots, v_k}$ and its elements by $g_{R; v_1, \dots, v_k}$. And for each $i\in [k]$, we identify the image of $\rel G^{v_i}$ under $e_{R,i}$ with its image in $\rel G_R^{v_1, \dots, v_k}$, i.e., the elements $g_{v_i}$ with $e_{R,i}(g)_{R; v_1, \dots, v_k}$ for each $g\in \rel G$.
  \end{itemize}
\end{definition}

It is well-known that such a gadget replacement can be computed in log-space.

\begin{example} \label{ex:arc-gadget}
  Let us show a simple gadget replacement from digraphs to digraphs. We use the following gadgets where we use the notation $\rel P_n  = (\{0, \dots, n\}; \{(i, i+1) \mid i < n)$ for a path of length $n$: $\rel G = \rel P_1$, and $\rel G_E = \rel P_2$ with the homomorphisms $e_{E,i}$ that map $\rel P_1$ to the first and second edge of $\rel P_2$ for $i = 1,2$ respectively. Graphically, they are represented as
  
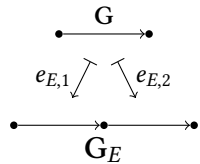
\[
    \begin{tikzpicture}[scale = 1.2]
      \node (x1) [circle, fill, inner sep=1pt] at (-.5, 1) {};
      \node (x2) [circle, fill, inner sep=1pt] at (.5, 1) {};
      \draw [->] (x1) -- (x2) node [midway, inner sep = 10] (p) {}
                              node [midway, above] {$\rel G$};

      \node (x1) [circle, fill, inner sep=1pt] at (-1, 0) {};
      \node (x2) [circle, fill, inner sep=1pt] at (0, 0) {}
        node [midway, below] {$\rel G_E$};
      \node (y2) [circle, fill, inner sep=1pt] at (1, 0) {};
      \draw [->] (x1) -- (x2) node [midway, inner sep = 10] (p1) {};
      \draw [->] (x2) -- (y2) node [midway, inner sep = 10] (p2) {};

      \draw [|->] (p) -- (p1) node [midway, left] {$e_{E,1}$};
      \draw [|->] (p) -- (p2) node [midway, right] {$e_{E,2}$};
    \end{tikzpicture}
  \]
  The gadget replacement defined by this gadget then replaces each vertex of the input graph with an edge, and if  $u$ is connected to $v$ by an edge, the endpoint of the new edge $u$ is identified with the starting point of $v$. For example, we get that $\phi(\rel P_1) = \rel P_2$, i.e., the path with two vertices becomes a path with two edges as we would expect, but also, more generally, $\phi(\rel P_n) = \rel P_{n+1}$ for all $n\geq 1$.
  See Fig.~\ref{fig:arc-gadget} for more examples of applying this gadget replacement.
  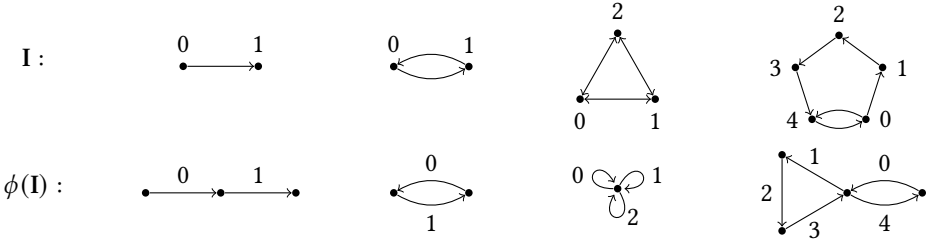
\begin{figure}
  \[
    \begin{matrix}
      \rel I : &\quad&
      \begin{tikzpicture}[baseline=(x1)]
        \node (x1) [circle, fill, inner sep=1pt, label={above:0}] at (-.5, 0) {};
        \node (x2) [circle, fill, inner sep=1pt, label={above:1}] at (.5, 0) {};
        \draw [->] (x1) -- (x2);
      \end{tikzpicture} &\quad&
      \begin{tikzpicture}[baseline=(x)]
        \node (x) [circle, fill, inner sep=1pt, label={above:0}] at (-.5, 1) {};
        \node (y) [circle, fill, inner sep=1pt, label={above:1}] at (.5, 1) {};
        \draw (y) edge [->, bend right] (x);
        \draw (x) edge [->, bend right] (y);
      \end{tikzpicture} &\quad&
      \begin{tikzpicture}[baseline=(current bounding box.center)]
        \node (x) [circle, fill, inner sep=1pt, label={below:0}] at (-.5, 0) {};
        \node (y) [circle, fill, inner sep=1pt, label={below:1}] at (.5, 0) {};
        \node (z) [circle, fill, inner sep=1pt, label={above:2}] at (0, .866) {};
        \draw (x) edge [<->] (y);
        \draw (y) edge [<->] (z);
        \draw (z) edge [<->] (x);
      \end{tikzpicture} &\quad&
      \begin{tikzpicture}[baseline=(current bounding box.center)]
        \node (0) [circle, fill, inner sep=1pt, label={right:0}] at (-54:.61) {};
        \node (1) [circle, fill, inner sep=1pt, label={right:1}] at (18:.61) {};
        \node (2) [circle, fill, inner sep=1pt, label={above:2}] at (90:.61) {};
        \node (3) [circle, fill, inner sep=1pt, label={left:3} ] at (162:.61) {};
        \node (4) [circle, fill, inner sep=1pt, label={left:4} ] at (234:.61) {};
        \draw (0) edge [->] (1);
        \draw (1) edge [->] (2);
        \draw (2) edge [->] (3);
        \draw (3) edge [->] (4);
        \draw (4) edge [->, bend right] (0);
        \draw (0) edge [->, bend right] (4);
      \end{tikzpicture} \\
      \phi(\rel I) : &\quad&
      \begin{tikzpicture}[baseline=(x1)]
        \node (x1) [circle, fill, inner sep=1pt] at (-1, 0) {};
        \node (x2) [circle, fill, inner sep=1pt] at (0, 0) {};
        \node (y2) [circle, fill, inner sep=1pt] at (1, 0) {};
        \draw [->] (x1) -- (x2) node [midway, above] {0};
        \draw [->] (x2) -- (y2) node [midway, above] {1};
      \end{tikzpicture} &\quad&
      \begin{tikzpicture}[baseline=(x)]
        \node (x) [circle, fill, inner sep=1pt] at (0, 0) {};
        \node (y) [circle, fill, inner sep=1pt]
          at (1, 0) {};
        \draw (y) edge [->, bend right] node [above] {0} (x);
        \draw (x) edge [->, bend right] node [below] {1} (y);
      \end{tikzpicture} &\quad&
      \begin{tikzpicture}[baseline=(current bounding box.center)]
        \node (x) [circle, fill, inner sep=1pt] at (0, 0) {};
        \draw (x) edge [in=0, out=60, loop] node [right] {1} (x);
        \draw (x) edge [in=120, out=180, loop] node [left] {0} (x);
        \draw (x) edge [in=240, out=300, loop] node [right] {2} (x);
      \end{tikzpicture} &\quad&
      \begin{tikzpicture}[baseline=(current bounding box.center)]
        \node (x) [circle, fill, inner sep=1pt] at (0, .5) {};
        \node (y) [circle, fill, inner sep=1pt] at (0, -.5) {};
        \node (z) [circle, fill, inner sep=1pt] at (.866, 0) {};
        \node (w) [circle, fill, inner sep=1pt] at (1.866, 0) {};
        \draw (x) edge [->] node [left] {2} (y);
        \draw (y) edge [->] node [below] {3} (z);
        \draw (z) edge [->] node [above] {1} (x);
        \draw (z) edge [->, bend right] node [below] {4} (w);
        \draw (w) edge [->, bend right] node [above] {0} (z);
      \end{tikzpicture}
    \end{matrix}
  \]
  \caption{An example of gadget replacement (Example~\ref{ex:arc-gadget})} \label{fig:arc-gadget}
  \end{figure}
\end{example}

In many previous papers, e.g., \cite{BBKO21, BKW17}, pp-powers as constructions on the templates have much more prominence than gadget replacements. We include them in this survey to show the connection between the two, and that they are in fact two different points of view on the same concept.

We define a \emph{pp-power} using the same data as a gadget reduction, i.e., structures $\rel G$, $\rel G_R$'s, and homomorphisms $e_{R,i}$'s. The corresponding pp-power $\rho$ of a structure $\rel B'$ similar to $\rel B$ (and hence to $\rel G$, and $\rel G_R$'s) is then defined as follows:
\begin{itemize}
  \item the domain of $\rho(\rel B')$ is the set $\hom(\rel G, \rel B')$ of all homomorphisms $p\colon \rel G \to \rel B'$, and
  \item a relation $R^{\rho(\rel B')}$ of arity $k$ is the set
    \[
      \{ (f \circ e_{R,1}, \dots, f \circ e_{R,k}) \mid f\colon \rel G_E \to \rel B' \}.
    \]
\end{itemize}
It is not hard to see that this definition gives (up to homomorphic equivalence\footnote{Two structures are \emph{homomorphically equivalent} if they are homomorphic to each other.}) the same structures as the pp-powers defined in \cite{BOP18, BKW17} using the language of logic.

The key relation between these two constructions is that for each $\rel I$ and $\rel B'$, we have that $\rel I \to \rho(\rel B')$ if and only if $\phi(\rel I) \to \rel B'$. This relation is called \emph{adjunction}, and it is enough to prove that $\phi$ is a reduction from $\CSP(\rho(\rel B))$ to $\CSP(\rel B)$, and also from $\PCSP(\rel A, \rho(\rel B))$ to $\PCSP(\phi(\rel A), \rel B))$ --- the proof is straightforward and can be found in \cite[Section 4.2]{KOWZ20}.

\subsubsection{Polymorphisms}
\label{sec:poly}

In this subsection, we define and explain some intuition about one of the key notions of the characterisation of gadget reductions: \emph{polymorphisms} of the template. Loosely speaking, a polymorphism is homomorphism from a \emph{power} (which we define below) of one structure to another (from one part of the template to the other).
An intuition why polymorphisms matter comes from the observation that whenever there is a valid gadget reduction from $\PCSP(\rel A, \rel A')$ to $\PCSP(\rel B, \rel B')$ then a specific gadget reduction that uses powers of $\rel B$ (which powers are used and how they interact depends on the structure $\rel A$) will give a reduction as well.

\begin{definition}
  The \emph{$n$-th power of $\rel B$}, denoted by $\rel B^n$, is a structure similar to $\rel B$ with the universe is $B^n$, the set of all $n$-ary tuples of elements of $\rel B$, and relations defined as follows.
  For each $k$-ary relation $R^{\rel B}$, the relation $R^{\rel B^n}$ is the set of all tuples of the form $(\tup r_1, \dots, \tup r_k)$ where $(\tup r_1(i), \dots, \tup r_k(i)) \in R^{\rel B}$ for each $i \in [n]$ (here, $\tup r(i)$ denotes the $i$-th entry of $\tup r$).
\end{definition}

Another way to describe the relations of the $n$-th power of $\rel B$ is: Consider all $k\times n$-matrices whose columns are tuples in $R^{\rel B}$. The relation $R^{\rel B^n}$ consists of all $k$-tuples of rows of such matrices (keeping the order, so each matrix corresponds to one tuple in $R^{\rel B^n}$), where each row is interpreted as an $n$-tuple of elements of $B$.

Polymorphisms are then simply functions that satisfy these power gadgets in the following sense.

\begin{definition}
  Let $(\rel B, \rel B')$ be a PCSP template.\footnote{We could also define polymorphisms between any pair of similar structures, but it is not hard to observe that there are none, unless there is a homomorphism between the two structures.} A \emph{polymorphism} from $\rel B$ to $\rel B'$ of arity $n$ is a homomorphism $f\colon \rel B^n \to \rel B'$. We also simply say that any such $f$ is a polymorphism of the template $(\rel B, \rel B')$.

  We denote the set of all such $n$-ary polymorphisms by $\Pol^{(n)}(\rel B, \rel B')$, and the set of all polymorphisms of arity $n \geq 1$ by $\Pol(\rel B, \rel B')$. Also, we write $\Pol(\rel B)$ for $\Pol(\rel B, \rel B)$.
\end{definition}

In other words, a function $f\colon B^n\rightarrow B'$ is an $n$-ary polymorphism from $\rel B$ to $\rel B'$ if, for every (say, $k$-ary) relation symbol $R$ the following holds: For every $k\times n$ matrix $M$ with entries from $B$ such that each column of $M$ is a tuple in $R^{\rel B}$, if we apply $f$ to each row of $M$ (treated as a tuple) then we obtain a column representing a tuple from $R^{\rel B'}$.

These polymorphisms naturally generalise polymorphisms of a single structure that were used in the algebraic approach to CSP \cite[see, e.g.,][Section 4]{BKW17}.

\begin{example}
  Recall Example~\ref{ex:approx-graph-col}. For any $n\ge 1$, the $n$-ary functions from $\Pol(\rel K_k,\rel K_c)$ are simply the $c$-colourings of $\rel K_k^n$, the $n$-th direct (or tensor) power of $\rel K_k$.
\end{example}

\subsubsection{Minors, minions, and minion homomorphisms}
\label{sec:min}

We would like to highlight the absence of algebras from this `algebraic approach to promise CSP', which makes this name a bit of a misnomer. The key property of algebras, or clones, that is missing in polymorphisms of PCSP templates is that the set $\Pol(\rel A,\rel B)$ is not closed under composition --- for example, if $f$ and $g$ are polymorphisms from $\rel A$ to $\rel B$ of arities $3$ and $2$ respectively, then the function $h$ defined by $h(x,y,w,z) = f(g(x,w),y,z)$ is not necessarily a polymorphism (as it would be if $\rel A=\rel B$). In general, the composition is not even well-defined! This means that, as far as we can see now, most of fundamental results from universal algebra do not apply in PCSP, and often there is no possibility of adapting such methods from the classical CSP. In PCSP, we have to work with much weaker structure on polymorphisms. This weaker structure is called a \emph{minion} and it is obtained from the fact that polymorphisms are always closed under taking \emph{minors}.

\begin{definition}
  An~$n$-ary function $f\colon A^n\to B$ is called a~\emph{minor} of an $m$-ary function $g\colon A^m \to B$ given by a~map $\pi \colon [m] \to [n]$ if
  \[
    f(x_1,\dots,x_n) = g(x_{\pi(1)},\dots,x_{\pi(m)})
  \]
  for all $x_1,\dots,x_n \in A$. We will use notation $g^\pi$ for $g(x_{\pi(1)},\dots,x_{\pi(m)})$ and write $f=g^\pi$.
\end{definition}

Alternatively, one can say that $f$ is a~minor of $g$ if it is obtained from $g$ by identifying variables, permuting variables, and introducing dummy variables. The classes of functions that are closed under this operations are called \emph{(function) minions}.

\begin{definition}
  A~\emph{function minion} $\clo M$ on a pair of sets $(A, B)$ is a~non-empty subset of $\{f\colon A^n\rightarrow B\mid n\ge 1\}$ that is closed under taking minors. For fixed $n\ge 1$, let $\clo M^{(n)}$ denote the set of $n$-ary functions from $\clo M$.\footnote{If $A = \emptyset$, we still require that functions in $\clo M$ have a well-defined arity, i.e., in this case $\clo M$ is an infinite set that contains one function of each arity.} Unless stated otherwise, we assume that all minions are defined on finite sets.
\end{definition}

It is very easy to check that $\Pol(\rel A,\rel B)$ is a minion for any PCSP template $(\rel A,\rel B)$.
One specific minion that is used much, especially in hardness proofs, is the minion $\clo P_A$ of all projections on a set $A$, where an $n$-ary $i$-th \emph{projection} (a.k.a.\ \emph{dictator}) on $A$ is the operation $p_n^i(x_1,\ldots, x_n)=x_i$. Let us denote the minion $\clo P_{\{0,1\}}$ simply by $\clo P$, it is easy to see that $\clo P$ is isomorphic to any minion $\clo P_A$ with $|A|\ge 2$.
It is also well-known \cite[see, e.g.,][]{BKW17} and easy to check directly that $\clo P=\Pol(\rel A)$ where $\rel A$ is the structure such that $\CSP(\rel A)$ is 3-{\Sat} (i.e., $A=\{0,1\}$ and the relations $R_i^{\rel A}$, $0\le i\le 3$, are defined by 3-clauses with $i$ negated variables, respectively).

We remark that each polymorphism minion  $\Pol(\rel A,\rel B)$ naturally carries a topological structure (see \cite{KOWZ20} for details). This is a potentially powerful tool  for investigating the complexity of PCSPs, but the study of this direction is still at its infancy.

The existence of gadget reductions can be characterised in terms of \emph{minion homomorphisms} between the polymorphism minions of the involved templates, we present this characterisation in the next subsection. A minion homomorphism is the natural notion of a structure preserving map between minions: the structure of a minion is given by the minor taking operations.

\begin{definition}\label{def:minor-hom}
  Let $\clo M$ and $\clo N$ be two minions (not necessarily on the same pairs of sets).  A~mapping $\xi\colon \clo M \to \clo N$ is called a~\emph{minion homomorphism} if
  \begin{enumerate}
    \item it preserves arities, i.e., $\ar(g)=\ar(\xi(g))$ for all $g\in \clo M$, and
    \item it preserves taking minors, i.e., for each $\pi\colon [m] \to [n]$ and each $g\in \clo M^{(m)}$ we have
    \[
      \xi(g)^\pi = \xi(g^\pi)
    .\]
  \end{enumerate}
\end{definition}

We note that item (1) is required in order for (2) to make sense, otherwise $\xi(g^\pi)$ and $\xi(g)$ couldn't form a minor using $\pi$.

\begin{example}
  Let us start with a trivial example. For each minion $\clo M$, there is a minion homomorphisms $\xi\colon \clo P \to \clo M$ defined as follows. Fix $h \in \clo M^{(1)}$ which exists since $\clo M$ is non-empty. And define $\xi(p_n^i)\colon x_1, \dots, x_n \mapsto h(x_i)$. Clearly, $\xi(p_n^i) \in \clo M^{(n)}$ since $\clo M$ is closed under taking minors. It is not hard to check that $\xi$ also preserves minor relations.
\end{example}

\begin{example}
\label{ex:k3k4}
 There is a~minion homomorphism from $\Pol(\rel K_3,\rel K_4)$ to the minion $\clo P$.
  This minion homomorphism is built on the following combinatorial statement proved in \cite[Lemma 3.4]{BG16}: for each (say, $n$-ary) $f\in \Pol(\rel K_3,\rel K_4)$, there exist $t\in K_4$
  (we will call any such $t$ a~\emph{trash colour}), a coordinate $i\in [n]$, and a map $\alpha\colon K_3 \to K_4$ such that
  \[
    f(a_1,\dots,a_n) \in \{ t, \alpha( a_i ) \}
  \]
  for all $a_1,\dots,a_n\in K_3$. In other words, if we erase the value $t$ from the table of $f$ then the remaining partial function depends only on $x_i$. Moreover, it is shown in \cite[Lemma 3.9]{BG16}, that while the trash colour $t$ is not necessarily unique, the coordinate $i$ is.
  We define $\xi\colon \Pol(\rel K_3,\rel K_4) \to \clo P$ by mapping each $n$-ary $f$ to $p_n^i$ for the $i$ that satisfies the above. One can check that this is indeed a minion homomorphism (see \cite[Example 2.22]{BBKO21} for more details).
\end{example}

\emph{Minor conditions} are systems of special functional equations (height-1 identities, in the language of universal algebra) that can be satisfied in a minion, and they are useful to show that there is no minion homomorphism from one minion to another. A minor condition is a collection of finitely many \emph{minor identities} which are formal expressions of the form:
\begin{equation} \label{id:minor}
  f(x_1, \dots, x_n) \equals g(x_{\pi(1)}, \dots, x_{\pi(m)})
\end{equation}
or $f \equals g^\pi$ for short. We use the symbol $\equals$ to emphasize that this is an equation rather than an equality of two functions. In particular, the variables in this equation are the function symbols (and the $x_i$'s are implicitly assumed to be universally quantified). We say that identity (\ref{id:minor}) is satisfied in a set of functions from $A$ to $B$ if this set has two specific functions $f$ and $g$ that make this equation an equality of functions, i.e., $f(x_1, \dots, x_n) = g(x_{\pi(1)}, \dots, x_{\pi(m)})$ for all $x_1, \dots, x_n \in A$. 

A minor condition is then said to be \emph{satisfied in a minion} $\clo M$ if for each function symbol appearing in it, there is a function in $\clo M$ of the corresponding arity, such that this collection of functions satisfy all the identities in the condition. It is easy to see that a minion homomorphism preserves satisfaction of minor conditions in the sense that if $\xi\colon \clo M \to \clo N$ is a minion homomorphism, then all minor conditions satisfied in $\clo M$ are also satisfied in $\clo N$. The converse is also true for polymorphism minions of finite PCSP templates, i.e., if every minor condition satisfied in $\clo M$ is also satisfied in $\clo N$ then there is a minion homomorphism from $\clo M$ to $\clo N$ \cite[Theorem 4.12(1)--(3)]{BBKO21}.

\begin{example}
  A commonly appearing minor condition is the following collection of three minor identities which is called \emph{(ternary) weak near unanimity}:
  \begin{align*}
    s(x, y) &\equals n(x, x, y) \\
    s(x, y) &\equals n(x, y, x) \\
    s(x, y) &\equals n(y, x, x)
  \end{align*}

  This minor condition is then satisfied in $\Pol(\rel K_3, \rel K_6)$. In order to show that we present $s\in \Pol^{(2)}(\rel K_3, \rel K_6)$ and $n \in \Pol^{(3)}(\rel K_3, \rel K_6)$: $s(x, y) = x$ for $x\in \{0, 1, 2\}$.
  \begin{align*}
    n(x, y, z) &= \begin{cases}
      a & \text{if $a$ appears at least twice among $x,y,z$, and}\\
      x + 3 & \text{otherwise}.
    \end{cases}
  \end{align*}
  It is not hard to check that both $s$ and $n$ are polymorphisms of the given template, and that they indeed satisfy all the identities above.

  We can also show that this minor condition is not satisfied in $\Pol(\rel H_2, \rel H_k)$ for any $k\geq 2$, where the structures $\rel H_k$ are from Example \ref{ex:hypergraph-colouring}. Indeed, if $s,n \in \Pol(\rel H_2, \rel H_k)$ would satisfy such a condition, we would get that
  \[
    n(0, 0, 1) = n(0, 1, 0) = n(1, 0, 0) = s(0, 1).
  \]
  which contradicts that $n$ is a polymorphism. To see this, consider the $3\times 3$ matrix whose columns $(0, 0, 1), (0, 1, 0), (1, 0, 0)$ are in the relation of  $\NAE_2$ . Applying $n$ to the three rows of this matrix, which are also $(0, 0, 1)$, $(0, 1, 0)$, $(1, 0, 0)$ results in the constant tuple $(s(0, 1), s(0, 1), s(0, 1))$ which is not in the relation of $\NAE_k$ for any $k$.

  Consequently, this shows that there is no minion homomorphism from $\Pol(\rel K_3, \rel K_6)$ to $\Pol(\rel H_2, \rel H_k)$ for any $k\geq 2$.
\end{example}

Some minor conditions are useless for the above purpose since they are satisfies in all minions (or, equivalently, in the minion $\clo P$ of projections), we call such conditions \emph{trivial}. An example of such condition would be a condition obtained from the ternary weak near unanimity by omitting any of the three identities. We remark that the problem of deciding triviality of minor condition is a different formulation of the \emph{Label Cover} problem. Moreover, every problem $\PCSP(\rel A,\rel B)$ is equivalent to the problem of distinguishing minor conditions that are trivial from those not satisfied in $\Pol(\rel A,\rel B)$. We refer to \cite[Section 3]{BBKO21} for a detailed exposition of the connection between PCSPs, minor conditions, and Label Cover.

Finally, let us briefly mention \emph{abstract minions} which are getting more and more popular in the literature \cite[e.g.,][]{BGWZ20,CZ22,CZ22a}.
In short, an  abstract minion is to a function minion what a group is to a permutation group. In long, abstract minion $\clo M$ is a collection of arbitrary non-empty sets $\{\clo M^{(n)}\mid n\ge 1\}$ with a well-defined \emph{minor-taking operations}, i.e., maps $(\cdot)^\pi\colon \clo M^{(n)} \to \clo M^{(m)}$ for each $\pi \colon [n] \to [m]$ that satisfy the (obvious) relations: $(g^\pi)^\sigma = g^{\sigma \circ \pi}$ and $g^{\id} = g$ if $\id$ is the identity. A minion homomorphism between such minions is defined in the same way as in Definition~\ref{def:minor-hom} with the difference that the symbols $g^\pi$ are interpreted as the minor taking operation of the corresponding minion applied to the corresponding element $g$. In the language of category theory, an abstract minion is the same as a functor from the category of non-empty finite sets to the category of non-empty sets, and a minion homomorphism is the same as a natural transformation between two functors.  We note that every abstract minion can be represented as a function minion on sets $(A, B)$ where $A$ is countable (and $B$ is also countable if $\clo M^{(n)}$ is countable for all~$n$).

\subsubsection{Characterisation of gadget reductions}

The following theorem is the result of a long line of gradual improvements and generalisations. Starting with \cite{JCG97} which introduced the notion of polymorphisms for CSPs, followed by \cite{BJK05,BOP18} which further refined the statement for CSPs, \cite{BBKO21} which generalised the scope to PCSPs, and \cite{KOWZ20} which mentioned the converse implication (1)$\to$(2) that has received very little attention before.

\begin{theorem} \label{thm:gadgets}
  Let $(\rel A, \rel A')$ and $(\rel B, \rel B')$ be two PCSP templates. Then the following two statements are equivalent:
  \begin{enumerate}
    \item $\PCSP(\rel A, \rel A')$ is reducible to $\PCSP(\rel B, \rel B')$ by a gadget reduction (and hence in log-space),
    \item there is a minion homomorphism $\Pol(\rel B, \rel B') \to \Pol(\rel A, \rel A')$.
  \end{enumerate}
\end{theorem}

Since polymorphisms of PCSP templates are not closed under composition, unlike polymorphisms of CSP templates, much of the structural theory of universal algebra is not applicable in the promise setting. On the other hand, since the above theorem is a direct generalisation of the corresponding result for CSPs \cite{BOP18}, every statement about PCSPs applies to CSPs as well, and therefore the study of PCSPs brings new insights into understanding CSPs as well.

One key concept that is useful for proving Theorem~\ref{thm:gadgets}, but also in different settings, is a notion of the \emph{free structure}. It stems from the observation that if we fix $\rel A$, $\rel B$, and $\rel B'$, there is always a structure $\rel A'$ such that $\PCSP(\rel A, \rel A')$ reduces to $\PCSP(\rel B, \rel B')$ by a gadget reduction.
Moreover, there is a universal such structure (i.e., it maps homomorphically to any suitable $\rel A'$), and this universal structure depends only on $\rel A$ and the polymorphisms from $\rel B$ to $\rel B'$. It is called the \emph{free structure of $\Pol(\rel B, \rel B')$ generated by $\rel A$} and denoted by $\rel F_{\Pol(\rel B, \rel B')}(\rel A)$ in \cite[Section 4.1]{BBKO21} to which we refer to for an explicit construction, formal definitions, and other useful properties \cite[see, in particular,][Theorem 4.12]{BBKO21}.

\begin{example}[3- vs 5-colouring]
  \label{ex:3vs5col}
  Recall the approximate (hyper)graph colouring problems from Examples \ref{ex:approx-graph-col} and \ref{ex:hypergraph-colouring}.
  There is a gadget reduction from hypergraph colouring $\PCSP(\rel H_2, \rel H_{27,480})$ to graph colouring $\PCSP(\rel K_3, \rel K_5)$.\footnote{The number $27,480$ is the number of binary polymorphisms from $\rel K_3$ to $\rel K_5$ which was given by a computer calculation.}
  This can be shown by arguing that every minor condition satisfied in $\Pol(\rel K_3, \rel K_5)$ is also satisfied in $\Pol(\rel H_2, \rel H_{27,480})$, and hence there is a minion homomorphism from $\Pol(\rel K_3, \rel K_5)$ to $\Pol(\rel H_2, \rel H_{27,480})$.

  Alternatively, using the claim in the previous paragraph, we could argue that $\PCSP(\rel H_2, \rel F)$ reduces to $\PCSP(\rel K_3, \rel K_5)$, where $\rel F$ is the free structure of $\Pol(\rel K_3, \rel K_5)$ generated by $\rel H_2$, and that $\rel F \to \rel H_{27,480}$ --- this is the essence of the proof presented in \cite[Section 6]{BBKO21}.

  One can also directly prove that the ``universal gadgets'' give a desired reduction from $\PCSP(\rel H_2, \rel H_{27,480})$ to graph colouring $\PCSP(\rel K_3, \rel K_5)$, see \cite[Section 4]{BKO19} that contains a self-contained (one-page) proof of this.
\end{example}

\section{``Rich enough'' polymorphisms, rounding, and tractability}
\label{sec:tract}

Currently, there is essentially one way to prove tractability of a problem $\PCSP(\rel A,\rel B)$, which goes as follows. Assume that we have a structure $\rel S$ (for sandwich) such that $\rel A\rightarrow \rel S\rightarrow \rel B$. It is easy to see that any algorithm solving $\CSP(\rel S)$ also solves the decision version of $\PCSP(\rel A,\rel B)$ --- indeed, for any input $\rel I$, if $\rel I\not\rightarrow \rel S$ then $\rel I\not\rightarrow \rel A$ and if $\rel I\rightarrow \rel S$ then $\rel I\rightarrow \rel B$.
We can even allow $\rel S$ to be an infinite structure (i.e., have an infinite domain) --- but the instances of $\rel S$ are still finite --- in fact, every instance of $\PCSP(\rel A,\rel B)$ is also an instance of $\CSP(\rel S)$. So, if this $\CSP(\rel S)$ is tractable, then so is $\PCSP(\rel A,\rel B)$. We remark, though, that many existing tractability proofs for PCSPs do not construct $\rel S$ explicitly. We will initially focus on decision PCSPs in this section, and briefly discuss search PCSPs at the end of the section.  

\begin{example}\label{ex1:sandwich}
  Consider $\PCSP(\rel T,\NAE_2)$ from Example~\ref{ex:1in3}. 
  Let $\rel S$ be the structure whose domain is $\mathbb{Z}$ and whose only relation is $\{(x,y,z)\in \mathbb{Z}^3 \mid x+y+z=1\}$. It is easy to see that $\rel T\rightarrow\rel S\rightarrow \NAE_2$. Indeed the first homomorphism is simply a (set-theoretic) inclusion, while the second one maps all negative integers to 0 and the rest to 1. The problem $\CSP(\rel S)$ amounts to solving linear systems over integers and thus belongs to $\Ptime$ \cite{KB79}. It follows that  $\PCSP(\rel T,\NAE_2)$ is in \Ptime. There are other choices for $\rel S$ that work for this particular PCSP.  
\end{example}

Some tractable problems $\PCSP(\rel A,\rel B)$ can be explained by the existence of a tractable sandwich CSP with a finite template $\rel S$, but the size of the smallest possible sandwich structure $\rel S$ can be arbitrarily large \cite{KMZ21} even when the sizes of domains of $\rel A$ and $\rel B$ are fixed.
However, there exist tractable PCSPs where tractable sandwich structures $\rel S$ exist, but are all necessarily infinite. One specific example is $\PCSP(\rel T,\NAE_2)$ from the example above --- it is shown in \cite[Section 8]{BBKO21} that $\CSP(\rel C)$ is \NP-complete for every finite structure $\rel C$ with $\rel T\rightarrow\rel C\rightarrow\NAE_2$.

It is an open question whether every tractable problem $\PCSP(\rel A,\rel B)$ sandwiches a tractable $\CSP(\rel S)$, possibly with infinite $\rel S$, and whether, for every $\PCSP(\rel A,\rel B)$ with this property, such a structure $\rel S$ can always be chosen to satisfy special nice properties. It is not even clear what such nice properties should be.

In all known cases, such an infinite sandwich structure $\rel S$ captures the possible outputs of solving a (polynomial-time solvable) relaxation of instances of $\CSP(\rel A)$, and then appropriate polymorphisms in $\Pol(\rel A,\rel B)$ provide a rounding procedure for this relaxation. The prime examples of this pattern are the basic linear programming (BLP) relaxation and the affine integer programming (AIP) relaxation, which we now introduce.

Let $\rel I$ be an instance of $\CSP(\rel A)$ and consider the following (standard) 0-1 integer linear program equivalent to this instance.
The variables in the program are $\mu_v(a)$ for every $v\in I$ and $a\in A$, and $\mu_{\tup v,R}(\tup a)$ for every relational symbol $R$, every $\tup v\in R^{\rel I}$ and $\tup a \in R^{\rel A}$. The intended meaning is that $\mu_v(a)=1$ if and only if $v$ is assigned $a$ and that
$\mu_{\tup v,R}(\tup a)=1$ if and only if $\tup v$ is assigned $\tup a$. We will write $\tup v(i)$ or $\tup a(i)$ for the $i$-th coordinate of the corresponding tuple. The equations in the program are 
  \begin{align}
    \sum_{a\in A} \mu_v(a) &= 1 & v\in I, \label{eq:lip1} \\
    \sum_{\tup a\in R^{\rel A}} \mu_{\tup v,R}(\tup a) &= 1 & \tup v\in R^{\rel I}, \label{eq:lip2} \\
    \sum_{\tup a \in R^{\rel A}, \tup a(i) = a} \mu_{\tup v,R}(\mathbf a) &= \mu_{\tup v(i)}(a) & a\in A, \tup v \in R^{\rel I}, i \in [\ar(R)]. \label{eq:lip3}
  \end{align}

If we relax the above program and allow all variables to take values in the interval $[0,1]$, we obtain an instance of the linear programming feasibility problem that we denote  by $\BLP_{\rel A}(\rel I)$. If we instead allow all variables to take arbitrary integer values, we obtain a system of linear equations over integers that we denote by $\AIP_{\rel A}(\rel I)$. It is known that both linear programming feasibility and systems of linear equations over integers can be solved in polynomial time.

It is obvious that if $\rel I\rightarrow \rel A$ then both $\BLP_{\rel A}(\rel I)$ and $\AIP_{\rel A}(\rel I)$ have solutions.  We say that BLP (respectively, AIP) solves $\PCSP(\rel A,\rel B)$ if we have $\rel I\rightarrow \rel B$ whenever $\BLP_{\rel A}(\rel I)$ (respectively, $\AIP_{\rel A}(\rel I)$) has a solution.  What are the PCSPs solvable by BLP or by AIP? We explain this in some detail for BLP, the case of AIP is similar.

  Given a solution to $\BLP_{\rel A}(\rel I)$, how can we round it to a homomorphism from $\rel I$ to $\rel B$? It is not hard to see that rational-valued solutions to $\BLP_{\rel A}(\rel I)$ precisely correspond to homomorphisms from $\rel I$ to an (infinite) structure $\rel S$, which depends on $\rel A$, constructed as follows. The domain of $\rel S$ is the set of all rational probability distributions over $A$, i.e., all solutions to the equation $\sum_{a\in A} x_a = 1$ (cf.\ equation  (\ref{eq:lip1})) in non-negative rational numbers. A tuple $\tup t$ of such probability distributions is in a relation $R^{\rel S}$ if there is a rational probability distribution $\gamma$ over $R^{\rel A}$ such that, for all $i$, the $i$-th marginal of $\gamma$ is the $i$-th component in $\tup t$ (cf.\ equations (\ref{eq:lip2}) and (\ref{eq:lip3})).
  The problem $\CSP(\rel S)$ is a subproblem of the LP feasibility problem and so is tractable. 
  To use this structure $\rel S$ to prove tractability of $\PCSP(\rel A,\rel B)$ for some $\rel B$, we need to have $\rel A\rightarrow \rel S\rightarrow \rel B$. It is easy to see that we have $\rel A\rightarrow\rel S$, by mapping each element in $A$ to the distribution assigning probability 1 to that element. For $n\ge 1$, let $\rel S_n$ be the (finite) substructure of $\rel S$ whose domain consists of all probability distributions where the probability of each element is $\ell/n$  for some $\ell\in \{0,1,\ldots, n\}$. By compactness, we have that $\rel S\rightarrow \rel B$ if and only if $\rel S_n\rightarrow \rel B$ for all $n$.
  It is not hard to see that homomorphisms from $\rel S_n$ to $\rel B$ are in 1-to-1 correspondence with $n$-ary polymorphisms $f\in \Pol(\rel A,\rel B)$ such that
  $f(x_1,\ldots, x_n)= f(x_{\pi(1)},\ldots,x_{\pi(n)})$ for all permutations $\pi$ of $[n]$ (and all values of the variables).
  Such functions are called \emph{symmetric}. To summarise this discussion, we have the following result (where we add condition (3) for completeness).
 
  \begin{theorem}[\cite{BBKO21}]
  \label{thm:blp}
  For any PCSP template $(\rel A,\rel B)$, the following are equivalent:
  \begin{enumerate}
      \item BLP solves $\PCSP(\rel A,\rel B)$,
      \item $\Pol(\rel A,\rel B)$ contains symmetric functions of all arities, and 
      \item there is a minion homomorphism from $\clo Q_\conv$ to $\Pol(\rel A,\rel B)$, where
      $\clo Q_\conv$ is the minion $\{\sum_{i=1}^n\alpha_ix_i\mid n\ge 1,\sum_{i=1}^n\alpha_i=1,  \alpha_i\in [0,1]\mbox{ for all } i\}$ of functions on $\mathbb Q$.
  \end{enumerate}
  \end{theorem}
 
  So, if $\BLP_{\rel A}(\rel I)$ has a solution then we have $\rel I\rightarrow\rel S$, and then, since $\rel I$ is finite,  $\rel I\rightarrow\rel S_n$ for some $n$ that is polynomially bounded in the size of $\rel I$. Then one can say that an $n$-ary symmetric polymorphism from $\rel A$ to $\rel B$ provides a rounding $\rel I\rightarrow\rel S_n\rightarrow \rel B$. Thus, if $\Pol(\rel A,\rel B)$ is ``rich enough'', here in the sense that it contains symmetric functions of all arities, then the decision version of $\PCSP(\rel A,\rel B)$ is in \Ptime.
  
  A similar reasoning works for AIP in place of BLP (in fact, Example~\ref{ex1:sandwich} can be seen as an application of AIP).  The "rounding" polymorphisms that naturally appear in this case have the following properties (think $x_1-x_2+x_3-\ldots-x_{2n}+x_{2n+1}$). Say that a function $f\colon A^{2n+1}\rightarrow B$ is \emph{alternating} if it is \emph{parity-symmetric}, that is,
  \begin{equation}\label{eq:bl-sym}
    f(x_1,\ldots, x_{2n+1}) = f(x_{\pi(1)},\ldots,x_{\pi(2n+1)})
  \end{equation}
  for all permutations $\pi$ of $[2n+1]$ that preserve parity (i.e., map odd numbers to odd and even to even), and additionally, $f$ satisfies the following \emph{cancellation} law: 
  \begin{equation}
    f(x_1,\ldots, x_{2n-1},y,y) = f(x_1,\ldots, x_{2n-1},z,z).
  \end{equation}

  \begin{theorem}[\cite{BBKO21}]
    \label{thm:aip}
  For any PCSP template $(\rel A,\rel B)$, the following are equivalent:
  \begin{enumerate}
      \item AIP solves $\PCSP(\rel A,\rel B)$,
      \item $\Pol(\rel A,\rel B)$ contains alternating functions of all odd arities, and 
      \item there is a minion homomorphism from $\clo Z_{\aff}$ to $\Pol(\rel A,\rel B)$, where
      $\clo Z_{\aff}$ is the minion $\{\sum_{i=1}^n\alpha_ix_i\mid n\ge 1,\sum_{i=1}^n\alpha_i=1, \alpha_i\in \mathbb Z\mbox{ for all } i\}$ of functions on $\mathbb Z$.
  \end{enumerate}
  \end{theorem}

  We remark that AIP was not studied in the dichotomy-led research on CSPs, so the potential of this algorithm as a subroutine in efficient algorithms for (P)CSPs and its relationship to other algorithms for CSPs \cite[e.g.,][]{BD06,IMMVW10} are currently not understood well.
 
  One particular way to combine BLP and AIP was considered in \cite{BGWZ20}. Roughly, this combined algorithm first solves $\BLP_{\rel A}(\rel I)$ and, if solvable, it finds a special solution $s$ for it (a relative interior point in the rational polytope of solutions). Then it strengthens $\AIP_{\rel A}(\rel I)$ by adding equations $x = 0$, where $x$ is any variable that is equal to 0 in $s$. Finally, it solves this modified $\AIP_{\rel A}(\rel I)$. The algorithm rejects if either of $\BLP_{\rel A}(\rel I)$ and the modified $\AIP_{\rel A}(\rel I)$ has no solution and accepts otherwise. This algorithm, called BLP+AIP, can also be seen as going via an appropriate sandwich structure $\rel S$, and the power of BLP+AIP  can also be characterised via polymorphisms. Several closely related characterisations were given in \cite{BGWZ20}, one of them is as follows.
 
 \begin{theorem}[\cite{BGWZ20}]
 \label{thm:blp+aip}
 BLP+AIP solves $\PCSP(\rel A,\rel B)$ if and only if $\Pol(\rel A,\rel B)$ contains parity-symmetric functions of all odd arities.
 \end{theorem}
 
 Theorem \ref{thm:blp+aip} also has an equivalent third condition similar to Theorems \ref{thm:blp} and \ref{thm:aip}. It involves an abstract minion $\clo M_{\conv+\aff}$ whose definition, though quite simple, is a bit too lengthy for this survey. See \cite{BGWZ20} for details.
 
 Using Theorem \ref{thm:blp+aip} and the description of tractable cases of Boolean CSPs \cite[see, e.g.,][]{BKW17}, it is easy to see that BLP+AIP solves all these tractable cases. However, there are simple examples of non-Boolean CSPs that are not solvable by this algorithm \cite[see][]{BGWZ20}. One can generalise BLP, AIP, and BLP+AIP to hierarchies such as the Sherali-Adams hierarchy for linear programming (see, e.g., \cite{CZ22,CZ22a} for some initial studies of applying such hierarchies for PCSPs). It is open whether using (some finite level of) such hierarchies gives an algorithm that can solve all tractable decision PCSPs (or even all tractable CSPs).
 
 Another interesting new aspect of PCSPs concerns the number of polymorphisms that are sufficient to guarantee tractability. 
It is known that the tractable CSPs can be characterised by the presence of a single polymorphism of arity 4 [see, e.g., \citealt{BKW17}], but all theorems in this section use infinite sequences of polymorphisms. This is not a coincidence, since in PCSP one cannot compose polymorphisms, and, in particular, one cannot produce useful polymorphisms of large arities from those of smaller arities. Moreover, it follows from \cite{BK22} that, for any finite family $F$ of (multi-variable) functions from $A$ to $B$, there is an \NP-hard problem $\PCSP(\rel A,\rel B)$ with $F\subseteq \Pol(\rel A,\rel B)$ (see also Corollary~\ref{cor:low-arity} and the discussion below it).
This appears to suggest that tractable PCSPs require infinite CSP algorithms, such as BLP or AIP.
The nature of polynomial-time algorithms that solve all tractable CSPs \cite{Bul17,Zhu17,Zhu20} is very finite. Does all this suggest that there are (potentially simpler) infinite-flavoured efficient algorithms for PCSPs, which, in particular, can solve all tractable CSPs?

 \subsection{Tractability of search PCSPs}
 
 Let us now discuss tractability for search PCSPs.
 When search is concerned, there is a difference between CSP and PCSP. As mentioned before, the search version of any tractable decision CSP is tractable, but this reduction from search to decision works only for PCSPs that sandwich a tractable $\CSP(\rel S)$ with finite $\rel S$. 
 Indeed, if $\rel A\rightarrow \rel S\rightarrow \rel B$ then, for any instance $\rel I$ of $\PCSP(\rel A, \rel B)$, we can solve $\rel I$ as an instance of the search problem for $\CSP(\rel S)$ and use a pre-computed homomorphism $\rel S \to \rel B$.
 However, when infinite sandwich structures $\rel S$ must be used, then we need to be able to efficiently compute both a homomorphism $h$ from an input structure $\rel I$ to $\rel S$ and a homomorphism from the substructure $h(\rel I)$ of $\rel S$ to $\rel B$ (i.e., both solve the relaxation and round the obtained solution). In this case, efficient search algorithms are currently known only for very few PCSPs, where the underlying (specific) polymorphisms can be easily evaluated to provide efficient rounding [see, e.g., \citealt{BG21}].
 
If we want to use (say, symmetric) polymorphisms as rounding for the search version of $\PCSP(\rel A, \rel B)$, we need to be able to evaluate them efficiently. More specifically, we need a sequence of polymorphisms $f_n\in\Pol(\rel A,\rel B)$ such that each $f_n$ is $n$-ary, and one can compute the value of $f_n$ on a given tuple in time polynomial in $n$.
One possible obstacle for this, though, is that the application of polymorphisms as rounding procedures for BLP+AIP in \cite{BGWZ20} uses polymorphisms whose arity can be exponential in the size of the instance.
It is currently not clear whether this obstacle is inherent (at least, for some cases) or can always be avoided by using polymorphisms in a suitable way. 
In general, however, the study of efficient polymorphisms, i.e., those that can be efficiently evaluated, is a new, wide open, and interesting direction. 

\subsection{Some open questions and directions}
\label{sec:tract-open}

To finish this section, let us mention several open questions and directions. While many of the questions and directions below are general, partial answers and new examples would be of considerable interest because we currently see only a small part of the complexity landscape in PCSP. This comment applies also to the lists of questions and directions in Sections \ref{sec:red-open} and \ref{sec:hard-open}.
\begin{itemize}
\item The tractable problem $\PCSP(\rel T,\rel H_2)$ from Examples \ref{ex:1in3} and \ref{ex1:sandwich} has the properties that both structures are Boolean (i.e., have domain $\{0,1\}$) and all their relations are symmetric (i.e., invariant under all permutations of coordinates). All tractable PCSPs with these properties have been described in \cite{BG21,FKOS19}, they all either admit a tractable finite sandwich or can be solved by BLP or AIP. Find other interesting examples of tractable Boolean (or indeed non-Boolean) PCSPs. 
\item Fix your favourite NP-hard problem of the form $\CSP(\rel A)$ and describe all tractable problems $\PCSP(\rel A,\rel B)$. (One can loosely call this describing tractable qualitative approximations to $\CSP(\rel A)$).
\item Find new interesting polynomial-time algorithms that can solve (some new) PCSPs. 
    \item Is it true that every tractable problem $\PCSP(\rel A,\rel B)$ sandwiches a (possibly infinite) tractable $\CSP(\rel S)$?
    \item Whenever $\PCSP(\rel A, \rel B)$ sandwiches a tractable $\CSP(\rel S)$, where $\rel S$ is infinite, can $\rel S$ always be chosen to satisfy special nice properties?
    \item What are the PCSPs that admit a finite sandwich structure with tractable CSP? (Such PCSPs are called \emph{finitely tractable} by \citet{AB21}).
    \item Are there other natural relaxations (and the corresponding structures and families of polymorphisms) that can be used to provide efficient algorithms for PCSPs? Can families of polymorphisms that play an important role in CSP (such as cyclic polymorphisms, \cite[see][]{BKW17}) suggest new relaxations (for which they can provide rounding)?
    \item 
    Can every tractable CSP (or even PCSP) be solved by combining BLP and AIP (possibly using some appropriate finite level of Sherali-Adams-like hierarchies)? 
    \item Is it true that the search version of $\PCSP(\rel A,\rel B)$ is tractable whenever the decision version is?  
    \item Is the search problem tractable for the PCSPs satisfying the conditions from Theorems \ref{thm:blp}--\ref{thm:blp+aip}?
    \item Can a problem $\PCSP(\rel A,\rel B)$ have an efficient search algorithm, even in the presence of symmetric polymorphisms of all arities, that is ``oblivious'' to the underlying polymorphisms? \cite{BGWZ20} 
    \item Is it true that when a search PCSP is tractable, this can always be explained by families of polymorphisms that can be evaluated in polynomial time?
    \item While we focus here mostly on the computational complexity of PCSPs, there was much research into descriptive complexity of CSPs, i.e., expressibility of problems $\CSP(\rel A)$ in various logics \cite[see, e.g.,][]{BKL08, ABD09}. Many questions in this direction are wide open for PCSPs.
    \item The two basic polynomial-time algorithms used to solve CSPs are 
    the bounded-width algorithm \cite{BK14} and the few subpowers algorithm \cite{BD06,IMMVW10}. Both use
    composition of polymorphisms  --- the former in its correctness proof and the latter in the actual algorithm.
    Are there ``composition-free'' algorithms that can replace them (i.e., solve the same CSPs)?
\end{itemize}

\section{Reductions that go beyond gadget replacements}

In the general algebraic theory of CSP, it was sufficient to group and compare CSPs by the gadget reductions (described in Section \ref{sec:gadgets}), so that two CSPs that are reducible to each other by gadget reductions are considered essentially the same. 
Moreover, it follows from the CSP Dichotomy Theorem \cite{Bul17,Zhu20} that (unless $\Ptime=\NP$) a CSP is NP-hard if and only if {3-\Sat} reduces to it via a gadget reduction. Of course, one can still use gadget reductions to reduce from {3-\Sat} (or any other NP-hard CSP) to PCSPs --- however,  there are NP-hard PCSPs whose NP-hardness provably cannot be shown by a gadget reduction from an NP-hard CSP. Examples of this include the $(2+\epsilon)$-{\Sat} problem from Example~\ref{ex:2+eps}, and all the state of the art results on approximate graph colouring from Example \ref{ex:approx-graph-col}.  Hardness arguments for these examples rely on ad hoc reductions from problems that are proven NP-hard by different methods, typically by reducing from the Gap Label Cover problem, which is NP-hard by the PCP theorem and the Parallel Repetition Theorem \cite{ALMSS98,AS98,Raz98}. This frequent use of ad hoc reductions was one reason why \cite{BBKO21} called for an extension of the algebraic approach with more reductions. Such reductions would provide a coarser grouping of PCSPs than the one provided by gadget replacement, where more PCSPs are considered essentially the same (than what is provided by Theorem \ref{thm:gadgets}). Obviously, this could expand the scope of tractability results. Moreover, we would like to have a general theory for PCSPs, where (as many as possible) hardness results are explained in a uniform way, ideally via reductions of some regular type (which ideally would be characterised in a similar way gadget reductions are characterised by polymorphisms) and from as few problems as possible (ideally just from {3-\Sat}). This section outlines a few of the recent advances in that direction.

A common feature of the new reductions described in this subsection is that they are, surprisingly, \emph{pp-powers} (recall Section~\ref{sec:gadgets-scope}) --- which is a construction that is efficiently computable but, when studying gadget reductions, is applied to the template rather than to an instance.

\subsection{The arc-graph reduction and adjunctions}
\label{sec:adj}

A reduction between PCSPs that provably goes beyond gadget replacements has been used in \cite{KOWZ20} to improve hardness results for approximate graph colouring. This reduction transforms a digraph into its \emph{arc-graph}, which is defined as follows.
For a digraph $\rel G$, its arc-graph $\delta(\rel G)$ is the digraph whose vertices are the edges of $\rel G$, and whose edges are pairs of edges of $\rel G$ forming a 2-path:
\[
  \delta(\rel G) = (
    E^{\rel G};
    \{ ((x, y), (z, w)) \mid (x, y), (z, w) \in E^{\rel G}, y = z \}).
\]
It is known that, for any digraphs $\rel G, \rel H$, we have 
$\delta(\rel G) \to \rel H$ if and only if  $\rel G \to \delta_R(\rel H)$, where
the transformation $\delta_R$ can be described in a few ways. For example, we can set $\delta_R(\rel H)$ to be the digraph whose vertices are all subsets of $H$, and there is an edge from a subset $U$ to a subset $V$ if there is $u\in U$ such that $(u,v) \in E^{\rel H}$ for all $v\in V$. It is not hard to check that $\delta$ is a reduction from 
$\PCSP(\rel G,\delta_R(\rel H)$ to $\PCSP(\delta(\rel G),\rel H)$ whenever both PCSPs are defined, i.e., when $\delta(\rel G) \to \rel H$ and $\rel G \to \delta_R(\rel H)$.

This reduction can be used to reduce from $\PCSP(\rel K_{b(k)}, \rel K_{b(c)})$ to $\PCSP(\rel K_k, \rel K_c)$ for any $k,c \geq 4$ where $b(k) = \binom{k}{\lfloor k/2 \rfloor}$. Starting with a result of \citet{Hua13}, a repeated application of this reduction has led to a considerable improvement in state-of-the-art NP-hardness results for approximate graph colouring \cite{KOWZ20}. We will discuss this in more detail in Section \ref{sec:hardGLC}.

The arc-graph reduction falls into a wider scheme of reductions that are described by \emph{adjunctions}, which is a notion from category theory. 
Two transformations $\gamma$ and $\omega$ are said to be \emph{adjoint} if, for all structures $\rel A$ and $\rel B$,
\begin{equation} \label{eq:adj}
  \gamma(\rel A) \to \rel B \text{ if and only if } \rel A \to \omega(\rel B),
\end{equation}
assuming that $\gamma$ transforms $\rel A$ to a structure similar to $\rel B$ and $\omega$ transforms $\rel B$ to a structure similar to $\rel A$.
In this case $\gamma$ is called a \emph{left adjoint} to $\omega$, and $\omega$ a \emph{right adjoint} to $\gamma$.
Another example of adjunction is when $\gamma$ is a gadget replacement, and $\omega$ the corresponding pp-power, as we described in Section \ref{sec:gadgets-scope}. As it is known from the algebraic approach, in this case $\gamma$ gives an efficiently computable reduction from $\CSP(\omega(\rel A))$ to $\CSP(\rel A)$ for each relational structure $\rel A$ (to which $\omega$ can be applied). This in fact generalises to arbitrary adjunctions (naturally, from the computational complexity perspective, we are only interested in cases where $\gamma$ is efficiently computable),
and to PCSPs: any efficiently computable left adjoint $\gamma$ to $\omega$ can be used as a reduction from $\PCSP(\rel B, \omega(\rel A'))$ to $\PCSP(\gamma(\rel B), \rel A')$ \cite{KOWZ20}. More information on the use of adjunctions in PCSP (and more examples) can be found in \cite{KOWZ20}.

We remark that it is currently  not known how (or whether) reductions given by adjunctions, even in the special case of the arc-graph reduction, can be characterised in terms of polymorphisms of the involved templates.

\subsection{Generalising minion homomorphisms}
  \label{sec:cgt}

We briefly outline the core results of a recent paper by \citet{BK22} that describes a reduction between PCSPs that is more general than gadget reductions and gives a sufficient condition for the existence of this reduction from $\PCSP(\rel A, \rel A')$ to $\PCSP(\rel B, \rel B')$ that is weaker than the existence of minion homomorphism from $\Pol(\rel B, \rel B')$ to $\Pol(\rel A, \rel A')$. Moreover, this sufficient condition is still expressed in terms of the two polymorphism minions. This result goes towards the general goal described in the beginning of this section. In particular, it allows one to to show NP-hardness of some PCSPs by a regular type of reduction from 3-{\Sat} through the use of polymorphisms. This covers several NP-hardness results, e.g., those from Examples \ref{ex:2+eps}, \ref{ex:approx-graph-hom}, and \ref{ex:hypergraph-colouring}, where original proofs relied on Gap Label Cover and the PCP theorem.

We formulate a simplified version of the reduction used in \cite{BK22} --- while formally, this formulation is different, it can be shown that if there is a reduction according to \cite{BK22} between two PCSPs, then the reduction described here also works.
We use the notation $\binom X{\leq k}$ for the set of all at most $k$-element non-empty subsets of $X$. And we refine a definition of a power of a structure to allow sets for exponents; the structure $\rel B^X$ is isomorphic to $\rel B^{\lvert X\rvert}$.

\begin{definition}
  Let $X$ be a set and $\rel B$ a relational structure, we define the \emph{$X$-th power of $\rel B$} denoted by $\rel B^X$ as the structure $(B^X; R^{\rel B^X}, \dots)$ where $B^X$ is the set of all functions from $X$ to $B$, and
  \[
    R^{\rel B^X} = \{ (b_1, \dots, b_k) \mid
      (b_1(x), \dots, b_k(x))\in R^{\rel B} \text{ for all } x\in X \}
  \]
  for each $k$-ary relational symbol $R$ of $\rel B$.
\end{definition}

We now give a concise description of the reduction which we will call here, for the lack of better name, \emph{$k$-reduction}. Assume that we are reducing from $\PCSP(\rel A, \rel A')$ to $\PCSP(\rel B, \rel B')$, and fix a (large enough) positive integer $k$.
Intuitively, the reduction is very similar to a gadget replacement with the key difference that we consider each set of at most $k$ variables at a time as one unit replaced by one gadget --- each of these gadgets is a power of $\rel B$ as defined above.

Let $\rel I$ be an instance of $\PCSP(\rel A, \rel A')$, we construct an instance $\phi(\rel I)$ as follows. For a subset $K\subseteq I$, let $\rel I[K]$ denote the (induced) substructure of $\rel I$ with domain $K$ obtained by removing all tuples containing elements not from $K$ from all relations in $\rel I$. 

\begin{itemize}
  \item For each $K \in \binom I{\leq k}$, let $\mathcal F_K$ be the set of all homomorphisms from $\rel I[K]$ to $\rel A$. For each such $K$, introduce into $\phi(\rel I)$ a copy of $\rel B^{\mathcal F_K}$.
  \item For each $L \subset K \in \binom I{\leq k}$, there is a natural map from $\mathcal F_K$ to $\mathcal F_L$ that maps $h$ to its restriction $h|_L$. We identify each element $b$ of $\rel B^{\mathcal F_L}$ with the  element $b_K$ of $\rel B^{\mathcal F_K}$ defined as $b_K(h) = b(h|_L)$.
\end{itemize}
It is not hard to see that this transformation $\phi$ can be computed in log-space.
When is this transformation a reduction from $\PCSP(\rel A, \rel A')$ to $\PCSP(\rel B, \rel B')$? \citet{BK22} provide a partial answer to this question.
We also note that they only claim that their reduction is polynomial time since in order to provide a log-space reduction between search problems, we need to provide an efficient way to reconstruct a solution for $\rel I$ from a solution $\phi(\rel I)$. Theorem~\ref{thm:dr-homomorphism} below only asserts that if there is a solution to $\phi(\rel I)$ then there is also a solution to $\rel I$, but it is not obvious whether it can be computed in log-space.

We remark that $k$-reduction is also closely connected with Sherali-Adams (SA) hierarchy  \cite[see][]{SA90, CZ22a} in the following sense. First, SA can be essentially viewed as a reduction from a CSP to linear programming. Naturally, this reduction is different from the basic linear programming relaxation. Namely, the difference lies in the fact that the $k$-th level of SA considers $k$ variables at the same time in very much the same way as in the definition above. A difference from the above construction is in the step where the sets $\mathcal F_K$ are converted into a gadget ($\rel B^{\mathcal F_K}$ above): SA uses a more efficient gadget consisting of all probability distributions on $\mathcal F_K$. This consequently also alters the identification step (second item above).

\Citet{BK22} introduce a generalisation of minion homomorphisms that is a sufficient condition for the existence of a $k$-reduction for some $k$. The simpler version of the new homomorphism is the so-called \emph{$d$-homomorphism} which is a mapping that assigns to each function from $\Pol(\rel B, \rel B')$ a list of at most $d$ \emph{candidates} from $\Pol(\rel A, \rel A')$ of the same arity, such that, for each minor relation $f = g^\pi$, there is a choice of candidates, $f'$ from image of $f$ and $g'$ from the image of $g$, that satisfy the same minor relation $f' = (g')^\pi$. (The standard minion homomorphism is thus a 1-homomorphism.) This notion further generalises to $(d, r)$-homomorphisms defined below; a $d$-homomorphism is a $(d, 1)$-homomorphism. Loosely speaking, a $d$-homomorphism forces us to weakly satisfy each minor relation, while a $(1, r)$-homomorphism does not require that each minor relation is satisfied, but at least one that from each chain of $r$ minor relations. A precise definition that combines both these behaviours is given below.

\begin{definition}
  \label{def:dr-homomorphism}
  Fix a minion $\clo M$, a \emph{chain of minors} in $\clo M$ of length $r$ is a sequence $t_0, \dots, t_r \in \clo M$ together with functions $\pi_{i, j}$ for $0\le i<j\le r$, such that $t_j = t_i^{\pi_{i, j}}$ for all $i<j$ and $\pi_{i, k} = \pi_{i, j} \circ \pi_{j, k}$, for all $i < j < k$.

  Let $\clo M$ and $\clo N$ be two minions. A \emph{$(d, r)$-homomorphism} from $\clo M$ to $\clo N$ is a mapping $\xi$ that assigns to each element of $\clo M$ a non-empty subset of $\clo N$ of size at most $d$ such that
  \begin{enumerate}
    \item for all $f\in \clo M$, every $g\in \xi(f)$ has the same arity as $f$.
    \item for each chain of minors in $\clo M$ as above, there exist $i < j$ and $g_i \in \xi(t_i)$, $g_j \in \xi(t_j)$ such that $g_j = g_i^{\pi_{i, j}}$.
  \end{enumerate}
\end{definition}

Note that, in a chain of minors, the maps $\pi_{i, j}$ are determined by $\pi_{i, i+1}$ since all others are obtained by composition. So, one way to create such a chain is to start with $t_0$ and to iteratively take a minor $t_{i+1} = t_i^{\pi_{i, i+1}}$.

Often a $(d, r)$-homomorphism can be given by providing $r$ partial $d$-homomorphisms that cover the whole minion $\clo M$. More precisely, assume that $\clo M = \clo M_1 \cup \dots \cup \clo M_r$ where each $\clo M_i$ is not necessarily a minion, and that we have maps $\xi_i\colon \clo M_i \to \clo N$ that are \emph{partial $d$-homomorphisms}, i.e., they satisfy the conditions (1) and (2) in Definition~\ref{def:dr-homomorphism} for every chain of minors of length 1 that is contained in the same $\clo M_i$. Then there is a $(d, r)$-homomorphism $\xi\colon \clo M \to \clo N$ which is defined by taking the union of $\xi_i$'s and breaking ties arbitrarily whenever some of $\xi_i$'s disagree. This is a $(d, r)$-homomorphism, since, for every chain of minors $t_0, \dots, t_r$, there will be some $s$ such that some $t_i$ and $t_j$ belong to the same $\clo M_s$ and $\xi(t_i) = \xi_s(t_i)$ and $\xi(t_j) = \xi_s(t_j)$.

\begin{theorem}[{\cite[Theorem 5.1]{BK22}}]
  \label{thm:drhom-red} \label{thm:dr-homomorphism}
  Let $(\rel A, \rel A')$ and $(\rel B, \rel B')$ be two PCSP templates.  If there is a $(d, r)$-homomorphism from $\Pol(\rel B, \rel B')$ to $\Pol(\rel A, \rel A')$, then there exists $k$ such that $\PCSP(\rel A, \rel A')$ reduces to $\PCSP(\rel B, \rel B')$ by the $k$-reduction.
\end{theorem}

One important consequence of the above theorem (Corollary \ref{cor:low-arity} below) is that polymorphisms of a bounded arity (or, more generally, of bounded essential arity) have no influence on the complexity of a PCSP.  In particular, one cannot make any conclusions about the complexity of $\PCSP(\rel A,\rel A')$ solely on the basis of the presence of any single finite set of functions in the polymorphism minion $\Pol(\rel A,\rel A')$.
A coordinate $i$ of a function $f\colon A^n\rightarrow A'$ is called \emph{inessential} if, for all $i\in [n]$ and $a_1, \dots, a_n, a_i' \in A$, we have $f(a_1,\dots,a_i,\dots,a_n)=f(a_1,\dots,a'_i,\dots,a_n)$. Otherwise, the coordinate $i$ is called \emph{essential}. The number of essential coordinates of a function is called its \emph{essential arity}.

\begin{corollary} \label{cor:low-arity}
  Fix any number $m\ge 1$. If there is a partial minion homomorphism
  \[
    \xi\colon \clo B \to \Pol(\rel A, \rel A')
  \]
  where $\clo B \subseteq \Pol(\rel B, \rel B')$ is the set of all polymorphisms of essential arity at least $m$ then $\PCSP(\rel A, \rel A')$ is reducible to $\PCSP(\rel B, \rel B')$ by $k$-reduction for some $k$.
\end{corollary}

By partial minion homomorphism in the above theorem, we mean that minors $f=g^\pi$ are preserved for all functions $f,g\in \clo B$.

As a special case of Corollary \ref{cor:low-arity}, consider the situation when $A=B$, $A'=B'$ and $\Pol(\rel A, \rel A')$ and $\Pol(\rel B, \rel B')$ have the same functions of essential arity at least $m$, for some $m$, (but there is assumption on functions of essential arity less than $m$ in the two minions). Then Corollary \ref{cor:low-arity} implies that $\PCSP(\rel A, \rel A')$ and $\PCSP(\rel B, \rel B')$ are reducible to each other. 

The proof of Corollary \ref{cor:low-arity} is straightforward, we can extend the partial homomorphism $\xi$ to a full $d$-homomorphism, where $d$ is the number of $m$-ary functions in $\Pol(\rel A, \rel A')$. To define the extension, map any $n$-ary function $f\in \Pol(\rel B, \rel B')$ of essential arity less then $m$ to the set of all $n$-ary functions in $\Pol(\rel A, \rel A')$ whose set of essential coordinates is contained in the set of essential coordinates of $f$. It is not hard to check that this indeed gives a $d$-homomorphism from $\Pol(\rel B, \rel B')$ to $\Pol(\rel A, \rel A')$, and one can apply Theorem \ref{thm:dr-homomorphism}.

An argument similar to the one before Theorem  \ref{thm:dr-homomorphism} gives the following.

\begin{corollary} \label{cor:typed-d-hom}
  Let $\clo A = \Pol(\rel A, \rel A')$ and $\clo B = \Pol(\rel B, \rel B')$
  If there is $r$ and a collection of $r$ partial minion $d$-homomorphism $\xi_i\colon \clo B_i \to \clo A$ where $\clo B_i \subseteq \clo B$ for each $i = 0, \dots, r-1$, and $\clo B = \bigcup_{i=0}^{r-1} \clo B_i$ then $\PCSP(\rel A, \rel A')$ is reducible to $\PCSP(\rel B, \rel B')$ by $k$-reduction for some $k$.
\end{corollary}

One open question relating $(d, r)$-homomorphisms with the arc-graph reduction is whether there is a~$(d, r)$-homomorphism for some $d$ and $r$ from $\Pol(\delta \rel G, \rel H)$ to $\Pol(\rel G, \delta_R\rel H)$ for any digraphs $\rel G$ and $\rel H$. We know that the arc-graph reduction is a reduction between the corresponding PCSPs. In fact, it is possible to show using the construction $\delta_R$ that $3$-reduction also provides such a reduction in this case. Nevertheless, it is not clear that the sufficient condition from Theorem~\ref{thm:dr-homomorphism} is satisfied in this case. If it is not, this would mean that $(d, r)$-homomorphisms are not necessary for $k$-reduction to work.

We are far from understanding reductions (beyond gadget reductions) in many ways: Is $k$-reduction sufficient to build a strong enough general theory for PCSPs? Is there a characterisation when a $k$-reduction is an actual reduction from one PCSP to another in terms of the polymorphism minions of the two templates? Would a more general reduction be more open to such a characterisation?

\subsection{Some open questions and directions}
\label{sec:red-open}
\begin{itemize}
  \item Find a class of reductions between PCSPs that is strong enough (say, to cover more than gadget reductions) together with a characterisation when such a reduction is applicable.
  \item Can reductions given by adjunctions be described in terms of polymorphisms? The special case of the arc-graph construction is a good place to start.
  \item Is $(d, r)$-homomorphism also a necessary condition for the existence of a $k$-reduction? If not, find a more general (polymorphism-based) condition that would be both sufficient and necessary for such reductions.
  \item Is there a~$(d, r)$-homomorphism for some $d$ and $r$ from $\Pol(\delta \rel G, \rel H)$ to $\Pol(\rel G, \delta_R\rel H)$ for all (or some interesting) digraphs $\rel G$ and $\rel H$?
  \item It is known \cite{KOWZ20} that each minion of polymorphisms carries a topological structure. Investigate how this topological structure can be used to provide more reductions between PCSPs.
\end{itemize}

\section{``Limited enough'' polymorphisms and NP-hardness}
\label{sec:hardness}

For CSPs, we know exactly which problems $\CSP(\rel B)$ are \NP-hard (assuming $\Ptime\ne \NP$), by \cite{Bul17,Zhu17,Zhu20}. In fact, the hardness part of the CSP Dichotomy Theorem is easy and has long been known \cite{BJK05}.  One way to characterise the \NP-hard CSPs is as follows \cite{BOP18}. Recall the minion $\clo P$ of all projections on $\{0,1\}$ from Section \ref{sec:min}. Then the NP-complete CSPs are the problems $\CSP(\rel B)$  such that there exists a minion homomorphism from $\Pol(\rel B)$ to $\clo P$. Since, as we mentioned in Section \ref{sec:min}, $\clo P$ is the polymorphism minion of the structure corresponding to 3-{\Sat}, this means that every \NP-hard $\CSP(\rel B)$ admits a gadget reduction from 3-{\Sat} (unless P = NP).
Since every projection of a given arity $n$ is uniquely determined by one of its coordinates, the existence of such a minion homomorphism means exactly that, for each function $f$ in $\Pol(\rel B)$, one can select one ``important'' coordinate $i_f$ in such a way that if $f = g^\pi$ then $i_f = \pi(i_g)$.

For PCSPs, the situation is more complicated, since, as mentioned before, there are many \NP-hard PCSPs whose polymorphism minion does not admits a minion homomorphism to $\clo P$ (and hence there is no gadget reduction from 3-\Sat). In fact, there is some belief \cite{BG21} that, unlike for CSPs, understanding hardness in PCSPs will be at least as difficult as understanding tractability (regardless of the question about the existence of dichotomy for PCSPs).

\subsection{Proving NP-hardness results via polymorphisms} 

A prevalent (so far) use of Theorem \ref{thm:drhom-red} to prove \NP-hardness of PCSPs is to show a reduction from 3-\Sat, i.e., with $\rel A = \rel A'$ being the template of 3-\Sat, and hence $\Pol(\rel A, \rel A') = \clo P$.

\begin{corollary}
\label{cor:drhom-hard}
   Let $(\rel B, \rel B')$ be a PCSP template.  If, for some fixed $d,r\ge 1$, there is a $(d,r)$-homomorphism from $\Pol(\rel B, \rel B')$ to $\clo P$ then $\PCSP(\rel B,\rel B')$ is \NP-hard.
\end{corollary}

Corollary~\ref{cor:drhom-hard} for the case $r = 1$ was proved in \cite{BBKO21} (the core of the argument appeared in \cite{AGH17}) and for the case $d = 1$ and arbitrary $r$ in \cite{BWZ21}, but both of these results predate \cite{BK22} and their original proofs relied on the PCP theorem. 

Let us spell out what the condition in the above corollary says. Fix a $(d,r)$-homomorphism $\xi\colon \Pol(\rel B, \rel B')\rightarrow \clo P$. If $f$ is an $n$-ary function in $\Pol(\rel B, \rel B')$ then $\xi(f)$ is a non-empty collection of (at most) $d$ projections $p_n^{s_1},\ldots, p_n^{s_d}$ of arity $n$, where $1\le s_j\le n$ for each $j$. 
In other words, $\xi$ can be viewed as selecting a set $\sel(f)=\{s_1,\ldots,s_d\} \subseteq [n]$  of (at most) $d$ ``special'' coordinates in $f$. Now assume that we have a chain of minors $t_0, \dots, t_r \in \Pol(\rel B, \rel B')$ such that $t_j = t_i^{\pi_{i,j}}$, for each $i<j$, where $\pi_{i, k} = \pi_{i, j} \circ \pi_{j, k}$, for all $i < j < k$. Then there must exist $i < j$ such that $\sel(t_j) \cap \pi_{i, j}(\sel(t_i))\ne\emptyset$. In other words, it cannot be that, for all $i,j$, the image under $\pi_{i,j}$ of the ``special'' coordinates in $t_i$ does not contain any ``special'' coordinates of $t_j$.

Informally, the fact that one can identify a bounded number of special coordinates in every polymorphism (in a coordinated way) shows that these polymorphisms are in some way ``lopsided'', and so $\Pol(\rel B, \rel B')$ is ``limited enough''.  To contrast this intuition with ``rich enough'' polymorphisms from Section \ref{sec:tract}, let us explain why there cannot be a $(d,r)$-homomorphism from $\Pol(\rel B, \rel B')$ to $\clo P$ if $\Pol(\rel B, \rel B')$ contains symmetric functions of all arities. Indeed, assume such a $(d,r)$-homomorphism $\xi$ exists for some $d,r$. Choose any symmetric function $t_0$ of arity $d\cdot (r+1)$ in $\Pol(\rel B, \rel B')$ and let $S=sel(t_0)$ be the set of (at most) $d$ special coordinates of $t_0$ selected by $\xi$. Choose a permutation $\pi$ on $[d\cdot (r+1)]$ so that $S$ intersects none of the sets $\pi(S),\pi^2(S),\ldots,\pi^{d(r+1)-1}(S)$, clearly this can be done. Consider the chain of minors $t_0, \dots, t_r$ where $t_i=t_0^{\pi^{i}}$ (and so $\pi_{i,j}=\pi^{j-i}$ for all $i<j$). Note that we have $t_0=t_1=\ldots=t_r$, since $t_0$ is symmetric, but $S=\sel(t_j)$ is disjoint from $\pi_{i, j}(\sel(t_i))=\pi^{j-i}(S)$ for all $i<j$, by the choice of $\pi$.

Let us now discuss how Corollary \ref{cor:drhom-hard} is applied. Naturally, it is all about selecting special sets of coordinates in polymorphisms. How exactly to select the special variables for a given application is the creative part.

Let's start with the case $r = 1$. In this case we want to select, for each polymorphism $f$, a set $\sel(f)$ of (at most) $d$ ``special'' coordinates in such a way that, for every minor relation $t_1 = t_0^\pi$ in $\Pol(\rel B, \rel B')$, the set of special coordinates in $t_0$, after applying $\pi$, i.e., $\pi(\sel(t_0))$, shares at least one element with the set $\sel(t_1)$. 

\begin{example}
  Assume that we can prove that each $f\in \Pol(\rel B, \rel B')$ has at least one and at most $d$ essential variables (which means that the values of the remaining, non-essential, variables have no effect on the output of $f$). 
  We can set $\sel(f)$ to be the set of essential variables of $f$.  In this case it is easy to check that if $t_1 = t_0^\pi$ then $\sel(t_1)$ and $\pi(\sel(t_0))$ can never be disjoint. This reasoning was used, for example, in \cite{AGH17} to prove \NP-hardness of $(2+\epsilon)$-{\Sat} problems (see Example~\ref{ex:2+eps}) --- though via the PCP theorem.
\end{example}

The case $r = 1$ can be applicable even when polymorphisms have an unbounded number of essential variables --- the special variables can be selected on the basis of some other property, as in Example \ref{ex:k3k4} or in the following example.

\begin{example}
\label{ex:topology}
Let  $\rel C_{2k+1}$ be an undirected cycle with $2k+1$ nodes, $k\ge 1$, and let $\rel K_3$ be a 3-clique. The approximate graph homomorphism problem $\PCSP(\rel C_{2k+1},\rel K_3)$ (recall Example \ref{ex:approx-graph-hom}) has been proved \NP-hard in \cite{KOWZ20} by applying Corollary \ref{cor:drhom-hard} as follows. 
One can associate a topological space to every graph. The space associated with an odd cycle is a circle, and the space associated with the $n$-th power of an odd cycle is an $n$-torus. Furthermore, any graph homomorphism $f$, and hence also a polymorphism, induces a continuous map $\hat f$ between the corresponding spaces, hence we assign to a polymorphism $f\colon \rel C_{2k+1}^n \to \rel K_3$ a continuous map $\hat f$ from $n$-torus to the circle. Observing that the main topological invariant of a continuous map from a circle to itself is its degree (or winding number), one can then prove, roughly, that $\hat{f}$ will have a bounded number of coordinates with non-zero degree. These coordinates are then chosen to form the set $\sel(f)$, and one can prove that $\sel(t_1)\cap\pi(\sel(t_0))\ne\emptyset$ holds whenever $t_1=t_0^{\pi}$.
\end{example}

The previous example is a rare (so far) case when the analysis of polymorphisms in a hardness proof for PCSP is performed not by direct combinatorics, but instead by first mapping them by a minion homomorphism to a different abstract minion, where the behaviour of their images is easier to understand. It would be of significant interest to find more examples of hardness of proofs of this sort, where the target abstract minion can be of topological or of some other nature.

Now let us discuss applications of Corollary \ref{cor:drhom-hard} with $r>1$.
One useful pattern is provided in the following example, taken from \cite{BWZ21}, where it is applied to prove $\NP$-hardness of certain generalisations of $(2+\epsilon)$-{\Sat} to non-Boolean domains.

\begin{example}
  Assume that we can define the notion of a set of coordinates, called \emph{smug sets} in \cite{BWZ21}, for functions in $\Pol(\rel B, \rel B')$ so that the following conditions are satisfied:
  \begin{itemize}
      \item every $f\in\Pol(\rel B, \rel B')$ has a smug set consisting of at most $d$ coordinates;
      \item no function in $\Pol(\rel B, \rel B')$ can have more than $r$ pairwise disjoint smug sets;
      \item if $f=g^{\pi}$ for some $g,f\in \Pol(\rel B, \rel B')$ then the $\pi$-preimage of any smug set in $f$ is a smug set in $g$.
  \end{itemize}
  Then, for any $f\in\Pol(\rel B, \rel B')$, we can set $\sel(f)$ to be an arbitrary smug set of variables of size at most $d$ in it. This does not in general guarantee that $\sel(t_1)\cap\pi(\sel(t_0))\ne\emptyset$ whenever $t_1=t_0^{\pi}$. However, the last two conditions in the definition of a smug set imply that, for any chain of minors $t_0, \dots, t_r \in \Pol(\rel B, \rel B')$, it is impossible that $\sel(t_j) \cap \pi_{i, j}(\sel(t_i))=\emptyset$ for all $i<j$.
\end{example}

Another useful pattern of applying Corollary \ref{cor:drhom-hard} with $r>1$ is when
functions in $\Pol(\rel B, \rel B')$ can be shown to exhibit $r$ different types of behaviour, so that  $\Pol(\rel B, \rel B')$ can be divided into $r$ (not necessarily disjoint) types, where the definition of $\sel(f)$ depends on the type of a polymorphism, and one can 
show that $\sel(g)\cap\pi(\sel(f))\ne\emptyset$ holds whenever $g=f^{\pi}$ and $g$ and $f$ are of the same type. This can also be described as representing $\Pol(\rel B, \rel B')$ as $\Pol(\rel B, \rel B') = \clo M_1 \cup \dots \cup \clo M_r$, and providing partial $d$-homomorphisms $\xi_i\colon \clo M_i \to \clo P$ (see Definition \ref{def:dr-homomorphism} and the discussion after it).

\begin{example}
A few simple applications of the above pattern can be found in \cite{BBB21}, where it is applied to problems $\PCSP(\rel T,\rel B)$ (as in Example \ref{ex:1in3}) with $\rel B$ being a structure with domain $\{0,1,2\}$. For example, consider the case when the relation of $\rel B$ consists
of triples $(0,0,1),(0,0,2),(1,1,2)$ and all triples obtained from them by permuting coordinates. For a function $f:\{0,1\}^n\rightarrow \{0,1,2\}$, call a set $X$ of coordinates of $f$ a $2$-set if $f$ evaluates to $2$ whenever all variables in $X$ are equal to 1. Also, define $E(f)$ to be the set of all coordinates $i$ 
such that $f(\mathbf{x}_i)\ne 0$, where $\mathbf{x}_i$ is the tuple where 1 appears only in position $i$. It is shown in \cite{BBB21} that each $f\in \Pol(\rel T,\rel B)$ has at least one of the following properties:
$f$ has a 2-set $X$ of size at most 2 or the set $E(f)$ has size at most 5. Whichever case applies for each $f$, select the corresponding set, $X$ or $E(f)$, to be $\sel(f)$, breaking ties arbitrarily when there is any choice. 
Thus, we have $\Pol(\rel T,\rel B)=\clo M_1\cup \clo M_2$, where $\clo M_i$ consists 
all functions $f\in \Pol(\rel T,\rel B)$ such that $\sel(f)$ is defined by the first or by the second property above.
It is then shown that this provides partial $5$-homomorphisms from each $\clo M_i$ to $\clo P$, and hence a $(5,2)$-homomorphism from 
$\Pol(\rel T,\rel B)$ to $\clo P$.
\end{example}

\NP-hardness results for various Boolean PCSPs that can be obtained by using Corollary \ref{cor:drhom-hard} (even though the original proofs are not always presented that way)
can be found in \cite{BG21,BGS21,BZ21,FKOS19}.

\subsection{Hardness of PCSPs from Gap Label Cover}
\label{sec:hardGLC}

\NP-hardness of the Gap Label Cover problem is the starting point of many hardness proofs in inapproximability. Many hardness results for graph and hypergraph colouring  were obtained 
via ad hoc reductions from Gap Label Cover, and will discuss some of them here (see \cite[Section~5]{BBKO21} for more details).
We remark that one advantage of hardness proofs via Corollary \ref{cor:drhom-hard} (or similar), as compared with reductions from Gap Label Cover is that the reduction itself is not ad hoc, and thus is more suitable for building a general theory. 

We start with two examples where the original hardness proofs used Gap Label Cover,
but subsequently new proofs via Corollary \ref{cor:drhom-hard} were found.

\begin{example}
  Recall the hypergraph colouring problem $\PCSP(\rel H_k,\rel H_c)$ from Example \ref{ex:hypergraph-colouring}. It was proved in \cite{DRS05} that $\PCSP(\rel H_k,\rel H_c)$ is \NP-hard for all $2\le k\le c$, by using a reduction from a multi-layered version of Gap Label Cover. It was then shown in \cite[Section 5.3]{BBKO21} how the proof from \cite{DRS05} can be directly translated into the language of polymorphisms. \Citet{Wro22} refined this to provide a short combinatorial proof that there is a way to select special coordinates in functions from $\Pol(\rel H_2,\rel H_c)$ that gives a $(d,r)$-homomorphism from $\Pol(\rel H_2,\rel H_c)$ to $\clo P$ for appropriately chosen $d$ and $r$, so this hardness result can now also be explained by using Corollary \ref{cor:drhom-hard}. 
\end{example}

\begin{example}
  Recall Example \ref{ex:3vs5col}. For any $k\ge 3$, the problem $\PCSP(\rel K_k,\rel K_{2k-1})$ has been proved \NP-hard in \cite{BBKO21} by providing a minion homomorphism from $\Pol(\rel K_k,\rel K_{2k-1})$ to $\Pol(\rel H_2, \rel H_c)$, where $c$ is the number of binary functions in $\Pol(\rel K_k,\rel K_{2k-1})$. From the previous example, there is a $(d,r)$-homomorphism from $\Pol(\rel H_2,\rel H_c)$ to $\clo P$, for some $d,r$.  It is easy to check that a composition of a minion homomorphism with a $(d,r)$-homomorphism is again a $(d,r)$-homomorphism, so we conclude that there is a $(d,r)$-homomorphism from $\Pol(\rel K_k,\rel K_{2k-1})$ to $\clo P$.
\end{example}

Let us now discuss hardness results for approximate graph colouring that have not been explained via polymorphisms so far. Notably, all state-of-the-art \NP-hardness results for $\Pol(\rel K_k,\rel K_c)$ with $6\le k\le c$ are in this category.

\begin{example}
  The strongest known \NP-hardness results for $\PCSP(\rel K_k,\rel K_c)$ with $k = 3,4,5$ are those from the previous example, with $c = 2k-1$.  In \cite{Hua13}, it was proved, by a reduction from Gap Label Cover, that $\PCSP(\rel K_k,\rel K_c)$ is \NP-hard for $c = 2^{\Omega(k^{1/3})}$ and $k$ large enough.  By starting from Huang's result and repeatedly applying the arc-graph reduction (see Section \ref{sec:adj}), it was shown in \cite{KOWZ20} that $\PCSP(\rel K_k,\rel K_c)$ is \NP-hard for all $k\ge 4$ and $c = \binom{k}{\lfloor k/2\rfloor} - 1$ (which coincides with $2k-1$ for $k=5$ and is the strongest known result for $k\ge 6$).
  We note that neither Huang's result nor the arc-graph reduction currently have an explanation in terms of polymorphisms.
  
   Another result from \cite{KOWZ20} obtained by using the arc-digraph construction is that it is enough to prove \NP-hardness of $\PCSP(\rel K_k, \rel K_c)$ for some fixed $k\ge 3$ and all $c\ge k$, the same will then follow for all $3\le k\le c$. 
\end{example}

Other examples of NP-hardness results that were obtained by ad hoc reductions from Gap Label Cover (and cannot currently be explained
in terms of polymorphisms) include results on rainbow vs.\ normal and strong vs.\ normal hypergraph colouring, see references in Example \ref{ex:hypergraph-colouring}.

\subsection{Some open questions and directions}
\label{sec:hard-open}
\begin{itemize}
    \item At the moment, Corollary \ref{cor:drhom-hard} is the most general known formalisation of what it means for a PCSP to have ``limited enough'' polymorphisms (to guarantee NP-hardness). Find other (incomparable or more general) formalisations.
    \item We mentioned symmetric Boolean PCSPs in Section \ref{sec:tract-open}. All such PCSPs that are not tractable are NP-hard \cite{BG21,FKOS19}. Find other interesting examples of NP-hard Boolean (or non-Boolean) PCSPs.
    \item Which of the problems $\PCSP(\rel T,\rel B)$ from Example \ref{ex:1in3} are NP-hard? Specifically, solve this for the case $\rel B=\rel {LO}_k$  whose domain is $\{0,1,\ldots, k-1\}, k\ge 3$, and the relation consists of all tuples $(x,y,z)$ that have a unique maximal element (or, in other words, if two of $x,y,z$ are equal then the third element is larger than the two equal ones). See \cite{BBB21,NZ22} for related results.
    \item Pick your favourite structure $\rel A$ with NP-hard $\CSP(\rel A)$ and describe all NP-hard problems $\PCSP(\rel A,\rel B)$. More generally, classify the complexity of $\PCSP(\rel A, \rel B)$ depending on $\rel B$.
     \item Recall that a smooth digraph is one that has neither sources nor sinks (in other words, each vertex has positive in- and out-degrees).
    See \cite{BKW17,BKN09} for an explicit description of smooth digraphs $\rel G$ with tractable (NP-hard, resp.) $\CSP(\rel G)$.
    Generalising the Brakensiek-Guruswami conjecture from Example \ref{ex:approx-graph-hom}, are all problems $\PCSP(\rel G, \rel H)$ NP-hard when $\rel G$ and $\rel H$ are smooth digraphs with NP-hard CSPs?
    \item Further develop the topology-based approach mentioned in Example \ref{ex:topology} to attack the approximate graph colouring problem (and possibly other PCSPs such as variants of hypergraph colouring mentioned in Example \ref{ex:hypergraph-colouring}).
    \item Topology has been applied in hardness proofs for PCSPs in different ways: combinatorial analysis via Borsuk-Ulam-style theorems \cite{ABP20,DRS05} and topological analysis of polymorphisms \cite{KOWZ20}. Investigate whether there is a common pattern behind these applications.
    \item In the introduction, we mentioned that the truth of some NP-hardness conjectures (related to variants of unique games) is known to imply
    NP-hardness of all approximate graph colouring problems. Are the reverse implications true for approximate graph colouring (or some other PCSPs)?
    \item Investigate logical inexpressibility for PCSPs (see \cite{AD22} for the first example of work in this direction).
\end{itemize}

\bibliographystyle{ACM-Reference-Format}

\end{document}